\documentclass[aps,prd,twocolumn,amsmath,amssymb,groupedaddress,nofootinbib]{revtex4-2}

\usepackage{graphicx}
\usepackage{dcolumn}
\usepackage{bm}
\usepackage{upgreek}
\usepackage{makecell}
\usepackage{hyperref}

\begin{document}


\title{Postprocessing of tilt-to-length noise with coefficient drifts in TianQin using a null time-delay interferometry channel}


\author{Zhizhao~Wang}
\affiliation{MOE Key Laboratory of TianQin Mission, TianQin Research Center for Gravitational Physics $\&$ School of Physics and Astronomy, Frontiers
Science Center for TianQin, Gravitational Wave Research Center of CNSA, Sun Yat-sen University (Zhuhai Campus), Zhuhai 519082, China.}

\author{Shuju~Yang}
\affiliation{MOE Key Laboratory of TianQin Mission, TianQin Research Center for Gravitational Physics $\&$ School of Physics and Astronomy, Frontiers
Science Center for TianQin, Gravitational Wave Research Center of CNSA, Sun Yat-sen University (Zhuhai Campus), Zhuhai 519082, China.}

\author{Kaihang~Wu}
\affiliation{MOE Key Laboratory of TianQin Mission, TianQin Research Center for Gravitational Physics $\&$ School of Physics and Astronomy, Frontiers
Science Center for TianQin, Gravitational Wave Research Center of CNSA, Sun Yat-sen University (Zhuhai Campus), Zhuhai 519082, China.}

\author{Xiaojie~Wang}
\affiliation{MOE Key Laboratory of TianQin Mission, TianQin Research Center for Gravitational Physics $\&$ School of Physics and Astronomy, Frontiers
Science Center for TianQin, Gravitational Wave Research Center of CNSA, Sun Yat-sen University (Zhuhai Campus), Zhuhai 519082, China.}

\author{Huizong~Duan}
\email[]{duanhz3@mail.sysu.edu.cn}
\affiliation{MOE Key Laboratory of TianQin Mission, TianQin Research Center for Gravitational Physics $\&$ School of Physics and Astronomy, Frontiers
Science Center for TianQin, Gravitational Wave Research Center of CNSA, Sun Yat-sen University (Zhuhai Campus), Zhuhai 519082, China.}

\author{Yurong~Liang}
\affiliation{MOE Key Laboratory of Fundamental Physical Quantities Measurement, Hubei Key Laboratory of Gravitation and Quantum Physics, PGMF, and School of Physics, Huazhong University of Science and Technology, Wuhan 430074, China}

\author{Xuefeng~Zhang}
\email[]{zhangxf38@mail.sysu.edu.cn}
\affiliation{MOE Key Laboratory of TianQin Mission, TianQin Research Center for Gravitational Physics $\&$ School of Physics and Astronomy, Frontiers
Science Center for TianQin, Gravitational Wave Research Center of CNSA, Sun Yat-sen University (Zhuhai Campus), Zhuhai 519082, China.}

\author{Hsien-Chi~Yeh}
\affiliation{MOE Key Laboratory of TianQin Mission, TianQin Research Center for Gravitational Physics $\&$ School of Physics and Astronomy, Frontiers
Science Center for TianQin, Gravitational Wave Research Center of CNSA, Sun Yat-sen University (Zhuhai Campus), Zhuhai 519082, China.}


\date{\today}

\begin{abstract}
Tilt-to-length (TTL) coupling is expected to be one of the major noise sources in the interferometric phase readouts in TianQin mission. Arising from the angular motion of spacecraft (SC) and the onboard movable optical subassemblies (MOSAs), TTL noise needs to be removed in postprocessing after suppressing the laser phase noise with time-delay interferometry (TDI) technique. In this article, we show that we can estimate the TTL coupling coefficients using the null TDI channel $\zeta$ and remove the TTL noise in the commonly used Michelson variables with the estimated coefficients. We introduce the theoretical model of TTL noise in TDI and consider linear drifts in the linear TTL coefficients for noise estimation and subtraction. The TTL coefficients with drifts are estimated successfully with an accuracy of $10\ \upmu\text{m/rad}$ in our numerical simulation. We discuss the impact of point-ahead angle compensation error and wavefront error, and find it necessary to estimate linear drift coefficients and quadratic TTL coefficients to keep TTL noise residuals below the 0.3 pm noise reference curve. However, the estimation accuracy suffers greatly from the correlation between yaw jitter measurements that contain the same SC jitter. Assuming all angular jitters induced by MOSAs are independent, choosing a frequency range with relatively higher MOSA yaw jitter noise levels is beneficial to the TTL coefficient estimation.
\end{abstract}

\maketitle

\section{\label{sec1}Introduction}

TianQin is a space-based observatory planned to be launched in the 2030s, aiming to detect gravitational waves (GWs) within a frequency range from 0.1 mHz to 1 Hz~\cite{TianQin}. As a quasi-equilateral triangular constellation consisting of three spacecraft (SC), TianQin requires high-precision measurements of displacement variations between test masses (TMs) at the picometer level. The armlengths of TianQin are roughly $1.73\times10^5\ \text{km}$, for which the detector will greatly suffer from laser frequency noise. By means of the time-delay interferometry (TDI) algorithm~\cite{TDI2021}, the overwhelming laser frequency noise can be suppressed in postprocessing assuming we have accurate knowledge of the armlengths.

Apart from the laser frequency noise, tilt-to-length (TTL) coupling is also a significant limiting factor in TianQin. TTL coupling is induced by angular jitter as well as misalignments in the optical system. It is often effectively characterized by several linear coupling coefficients with the unit of mm/rad. It is reported that the expected level of TTL coupling coefficients after integration is 8.5 mm/rad in LISA, a space mission for GW detection led by the European Space Agency (ESA)~\cite{LISA2024}, and can be partly compensated to within 2.3 mm/rad by the so-called beam alignment mechanism (BAM)~\cite{Paczkowski2022}. However, the expected TTL noise after compensation still exceeds the noise requirement, which needs to be subtracted in postprocessing. As for TianQin, we expect the effective TTL contribution will not exceed 3 mm/rad after beam realignment, and the TTL noise allocation after postprocessing is $0.3\ \text{pm/Hz}^{1/2}$. Assuming the angular jitters are $10\ \text{nrad/Hz}^{1/2}$, the accuracy of the estimated TTL coefficients in postprocessing should be within $10\ \upmu\text{m/rad}$, which is a great challenge for TianQin.

\begin{figure}[htbp]
\includegraphics[width=8.6cm]{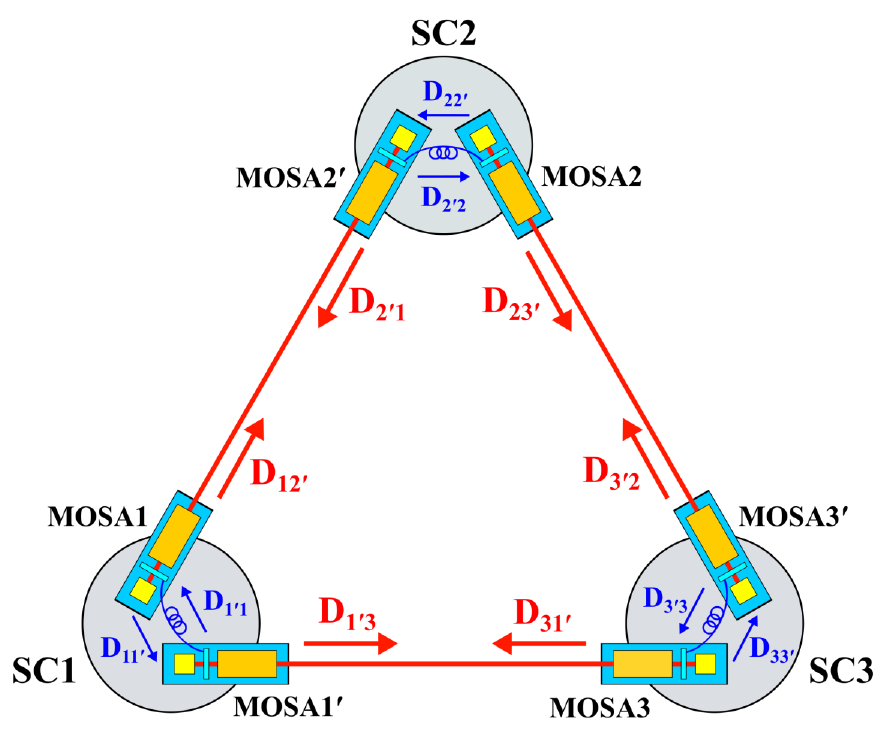}
\caption{\label{TQ_constellation} Labeling conventions used for spacecraft, MOSAs and laser links. The test mass, optical bench and telescope in each MOSA are drawn in yellow, cyan and orange respectively. The red lines represent the laser links while the blue curves between optical benches represent the backlink fibers for TMIs and RFIs.}
\end{figure}

As shown in Fig.~\ref{TQ_constellation}, there will be two movable optical subassemblies (MOSAs) on each TianQin SC. Each MOSA will house a free-falling TM, an optical bench (OB) and a telescope. The OB will integrate at least three types of interferometers: interspacecraft interferometer (ISI), test mass interferometer (TMI) and reference interferometer (RFI), whose readouts will be used to formulate the TDI variables. In postprocessing, one can combine the ISI and TMI readouts of each single link to remove the SC translational jitter along the sensitive axis, and cancel the TTL error that occurs in both ISI and TMI with opposite signs~\cite{Wanner2024}. The remaining TTL error in ISI and TMI should be within 3 mm/rad. TTL error in RFI is not considered as we assume there are no jittering components in RFI. As the TM motion is assumed to be slow, the expected jittering components at high frequencies are the SC and MOSAs. Therefore, we assume the TTL coupling effects in ISI and TMI at high frequencies share the same set of angular jitters.

Each interferometric signal is received by a quadrant photoreceiver and finally extracted by four independent digital phase-locked loops (DPLLs). The final interferometric readout is computed as the average of the four phase outputs from the DPLLs, while the angular misalignment between wavefronts of interfering beams is obtained by computing the differential phase, which is known as the differential wavefront sensing (DWS) technique~\cite{Morrison1994}. The ISI DWS readouts are considered as the total angular jitters induced by the SC and MOSA attitude control with respect to the incoming beam, which are used for TTL estimation in postprocessing.

In space missions like LISA Pathfinder and GRACE Follow-On, TTL coupling coefficients are estimated by performing in-orbit calibration maneuvers~\cite{Armano2023,Wegener2020}. Studies on TTL calibration maneuvers for LISA can be found in~\cite{Houba2022optimal,Houba2022}. However, it has been investigated in~\cite{Paczkowski2022,George2023} that TTL noise can be calibrated without disturbing the science operations. Among the 210 TDI combinations introduced in~\cite{Muratore2022}, the second-generation Michelson combinations $X$, $Y$ and $Z$ centered around SC1, SC2 and SC3 resepectively, are preferred for TDI variable construction since they are sensitive to GWs. Consequently, TTL coefficients is usually estimated after propagating the phase and angle measurements through $X$, $Y$ and $Z$ in previous studies, which is inevitably affected by GWs. To solve this problem, one can iteratively perform a global fit including all GW contents and TTL noise. As an alternative, we will introduce an approach for TTL estimation using null TDI channels~\cite{Muratore2022,Muratore2023} which allows significant suppression of GW signals. The typical null channel $C_3^{12}$, which is referred to as $\zeta$, is used for TTL estimation in this paper.

In~\cite{Paczkowski2022} each TTL coefficient are simulated with a linear drift of 0.15 mm/rad/day, while the authors in~\cite{George2023} assume the linear drifts are in the order of $20\ \upmu\text{m/rad/day}$. However, the coefficient drifts are not considered in TTL subtraction in both studies. Since the noise requirement for TianQin is more stringent than LISA, the impact of drifts may be non-negligible. In this paper, we will consider linear drifts in TTL coupling coefficients during parameter estimation. As a result, the impact of coefficient drifts will be removed after TTL noise subtraction.

This paper is organized as follows. We briefly introduce the postprocessing scheme of laser and clock noise removal in Sec.~\ref{sec2}, and give the TTL noise model in Sec.~\ref{sec3}. The numerical simulation setup and results are presented in Sec.~\ref{sec4} and Sec.~\ref{sec5} respectively. In Sec.~\ref{sec6}, we disucss the influences of pointing error and MOSA yaw jitters. Finally, we conclude in Sec.~\ref{sec7}.

\section{\label{sec2}Laser and Clock Noise Suppression}

Fig.~\ref{TQ_constellation} shows the schematic of TianQin constellation and the labeling conventions used in this paper. For any signal $x(t)$, we define the delay operator ${\bf D}_{ji}$ as
\begin{equation}
	{\bf D}_{ji}x(t) = x(t - d_{ji}(t))
\end{equation}
where $d_{ji}$ denotes the light travel time between MOSA$j$ and MOSA$i$, with $i,j\in\{1,1',2,2',3,3'\}$, $i \neq j$. We also introduce the advancement operator ${\bf A}_{ji}$, the inverse of ${\bf D}_{ji}$, given by
\begin{equation}
	{\bf A}_{ji}x(t) = x(t + \bar{d}_{ji}(t))
\end{equation}
with $\bar{d}_{ji}(t) = d_{ji}(t + \bar{d}_{ji}(t))$. Both operators will be used in the following text.

There are six lasers on the three SC. One of the six lasers is frequency-stabilized as the primary laser while the others are frequency offset phase-locked to the primary laser. This ensures that all beatnote frequencies fall into the bandwidth of quadrant photoreceivers. As a result, each interferometric phase measurement contains a phase ramp, which can be removed by a high-pass filter together with other out-of-band errors.

The filtered carrier-to-carrier interferometric measurements of ISI, TMI and RFI can be written as
\begin{widetext}
	\begin{subequations}
		\label{carrier_meas}
		\begin{align}
			\text{isi}_i &= p_i - {\bf D}_{ji}p_{j} + h_{ji} + \frac{\nu_{j}}{c}\left(-\boldsymbol{e}_{ij}\cdot\boldsymbol{\Delta}_i-\boldsymbol{e}_{ji}\cdot {\bf D}_{ji}\boldsymbol{\Delta}_{j}\right) + n_{ji,\text{isi}}^{\text{TTL}} + n_{i,\text{isi}}^{\text{opt}} + n_{i,\text{isi}}^{\text{ro}} - a_i q_i, \label{carrier_meas-a}\\
			\text{tmi}_i &= p_i - p_{i'} - \frac{2\nu_{i}}{c}\left(\boldsymbol{e}_{ij}\cdot\boldsymbol{\Delta}_i-\boldsymbol{e}_{ij}\cdot\boldsymbol{\delta}_i\right) - n_{i,\text{tmi}}^{\text{TTL}} + n_{i'i}^{\text{bl}} + n_{i,\text{tmi}}^{\text{opt}} + n_{i,\text{tmi}}^{\text{ro}}  - b_i q_i, \label{carrier_meas-b}\\
			\text{rfi}_i &= p_i - p_{i'} + n_{i'i}^{\text{bl}} + n_{i,\text{rfi}}^{\text{opt}} + n_{i,\text{rfi}}^{\text{ro}} - b_i q_i, \label{carrier_meas-c}
		\end{align}
	\end{subequations}
with $i''=i$. Notations $c$, $\nu$, $p$, $q$, $h$, $\boldsymbol{\Delta}$, $\boldsymbol{\delta}$, $n^{\text{TTL}}$, $n^{\text{bl}}$, $n^{\text{opt}}$, $n^{\text{ro}}$ denote the speed of light in vacuum, laser frequency, laser phase noise, clock phase noise, GW signal, SC translational jitter, TM translational jitter, TTL noise, backlink fiber pathlength noise, optical polarization and stray light noise, readout noise, respectively. $\boldsymbol{e}_{ji}$ denotes the unit vector pointing from MOSA$j$ to MOSA$i$.

The filtered sideband-to-sideband interferometric measurements used for clock noise reduction can be written as
\begin{subequations}
	\label{sideband_meas}
	\begin{align}
		\text{isi}_i^{\text{sb}} &= p_i + k_i q_i + m_i - {\bf D}_{ji}(p_{j} + k_{j}q_{j} + m_{j}) + h_{ji}^{\text{sb}} + \frac{\nu_{j}+k_j f_j^{\text{u}}}{c}\left(-\boldsymbol{e}_{ij}\cdot\boldsymbol{\Delta}_i-\boldsymbol{e}_{ji}\cdot {\bf D}_{ji}\boldsymbol{\Delta}_{j}\right) + n_{ji,\text{isi}}^{\text{TTL,sb}} \nonumber\\ &\quad + n_{i,\text{isi}}^{\text{opt,sb}} + n_{i,\text{isi}}^{\text{ro,sb}} - a_i^{\text{sb}} q_i, \label{sideband_meas-a}\\
		\text{tmi}_i^{\text{sb}} &= p_i + k_i q_i + m_i - (p_{i'} + k_{i'}q_{i'} + m_{i'}) - \frac{2(\nu_{i}+k_i f_i^{\text{u}})}{c}\left(\boldsymbol{e}_{ij}\cdot\boldsymbol{\Delta}_i-\boldsymbol{e}_{ij}\cdot\boldsymbol{\delta}_i\right) - n_{i,\text{tmi}}^{\text{TTL,sb}} + n_{i'i}^{\text{bl,sb}} \nonumber\\ &\quad + n_{i,\text{tmi}}^{\text{opt,sb}} + n_{i,\text{tmi}}^{\text{ro,sb}} - b_i^{\text{sb}} q_i, \label{sideband_meas-b}\\
		\text{rfi}_i^{\text{sb}} &= p_i + k_i q_i + m_i - (p_{i'} + k_{i'}q_{i'} + m_{i'}) + n_{i'i}^{\text{bl,sb}} + n_{i,\text{rfi}}^{\text{opt,sb}} + n_{i,\text{rfi}}^{\text{ro,sb}} - b_i^{\text{sb}} q_i, \label{sideband_meas-c}
	\end{align}
\end{subequations}
with $i''=i$ and $q_{i'}=q_i$. Notations $m$ and $k$ denote the modulation noise and the clock multiplier factor respectively. We have $k=k_1=k_2=k_3>0$ and $k'=k_{1'}=k_{2'}=k_{3'}>0$ for the upper sideband measurements. Given the onboard ultrastable oscillator (USO) frequency $f_i^{\text{u}}$, the fractional frequency beatnote coefficients in Eqs.~(\ref{carrier_meas}) and (\ref{sideband_meas}) read
\begin{subequations}
	\label{coeff_ab}
	\begin{align}
		a_i &= \frac{\nu_i-(1-\dot{d}_{ji})\nu_{j}}{f_i^{\text{u}}},\ a_i^{\text{sb}} = \frac{(\nu_i+k_i f_i^{\text{u}})-(1-\dot{d}_{ji})(\nu_{j}+k_j f_j^{\text{u}})}{f_i^{\text{u}}}, \\
		b_i &= \frac{\nu_i-\nu_{i'}}{f_i^{\text{u}}},\ b_i^{\text{sb}} = \frac{(\nu_i+k_i f_i^{\text{u}})-(\nu_{i'}+k_{i'}f_{i'}^{\text{u}})}{f_i^{\text{u}}},
	\end{align}
\end{subequations}
with $i''=i$ and $f_{i'}^{\text{u}}=f_i^{\text{u}}$.
\end{widetext}

\subsection{Laser noise suppression}

The primary aim of TDI is to reduce the overwhelming laser phase noise by constructing a virtual interferometer. The first step is to remove the OB jitter along the sensitive axis induced by the SC and MOSA motion, i.e.,
\begin{equation}
	\xi_i = \text{isi}_i - \frac{1}{2}\left(\frac{\nu_j}{\nu_i}(\text{tmi}_i-\text{rfi}_i)+{\bf D}_{ji}(\text{tmi}_j-\text{rfi}_j)\right). \label{xi}
\end{equation}

The second step is to replace the primed laser phase noise with the unprimed one, e.g.,
\begin{subequations}
	\label{eta}
	\begin{align}
		\eta_1 &= \xi_1 - \frac{1}{2}{\bf D}_{2'1}(\text{rfi}_2-\text{rfi}_{2'}), \\
		\eta_{1'} &= \xi_{1'} - \frac{1}{2}(\text{rfi}_{1'}-\text{rfi}_1).
	\end{align}
\end{subequations}
By cyclic permutation of the indices, one can compute the remaining four $\eta$-terms.

The last step is to construct the TDI variable to remove the unprimed laser phase noise. For example, the second-generation Michelson combination $X$ reads
\begin{align}
	X &= (1-{\bf D}_{31'}{\bf D}_{1'3}{\bf D}_{2'1}{\bf D}_{12'})[(\eta_1+{\bf D}_{2'1}\eta_{2'}) \nonumber\\ &\quad + {\bf D}_{2'1}{\bf D}_{12'}(\eta_{1'}+{\bf D}_{31'}\eta_3)] \nonumber\\ &\quad - (1-{\bf D}_{2'1}{\bf D}_{12'}{\bf D}_{31'}{\bf D}_{1'3})[(\eta_{1'}+{\bf D}_{31'}\eta_3) \nonumber\\ &\quad + {\bf D}_{31'}{\bf D}_{1'3}(\eta_1+{\bf D}_{2'1}\eta_{2'}))]. \label{X}
\end{align}
The null channel $\zeta$ used in this article reads
\begin{align}
	\zeta &= ({\bf D}_{23'} - {\bf D}_{1'3}{\bf A}_{12'})[{\bf D}_{3'2}{\bf A}_{31'}(\eta_{1'}-\eta_1)-\eta_2] \nonumber\\ &\quad - ({\bf D}_{1'3}{\bf A}_{12'}(1-{\bf D}_{3'2}{\bf A}_{31'}{\bf D}_{2'1})\eta_{2'} \nonumber\\ &\quad + (1-{\bf D}_{23'}{\bf D}_{3'2}{\bf A}_{31'}{\bf D}_{2'1}{\bf A}_{23'})(\eta_3-\eta_{3'}).
\end{align}

\subsection{Clock noise suppression}

The USO timing jitter will couple into the interferometric measurements. This will violate the requirement, so it must be reduced~\cite{Hartwig2021}. The clock noise reduction algorithm is implemented after removing the laser phase noise. One can obtain the differential clock jitter used for clock noise removal with the sideband measurements.

We first subtract the carrier phase measurements from the upper sideband measurements, i.e.,
\begin{subequations}
	\label{modulation_meas}
	\begin{align}
		\text{isi}_i^{\text{m}} &= \text{isi}_i^{\text{sb}} - \text{isi}_i, \\
		\text{tmi}_i^{\text{m}} &= \text{tmi}_i^{\text{sb}} - \text{tmi}_i, \\
		\text{rfi}_i^{\text{m}} &= \text{rfi}_i^{\text{sb}} - \text{rfi}_i.
	\end{align}
\end{subequations}
One can also combine the ISI and TMI readouts to remove the residual OB jitter in $\text{isi}_i^{\text{m}}$ associated with the modulation frequency. Here we asuume this term is already negligible and thus we skip this procedure. As a result, we have
\begin{equation}
	\xi_i^{\text{m}} = \text{isi}_i^{\text{m}}.
\end{equation}

Then we can replace the primed modulation noise with the unprimed one, e.g.,
\begin{subequations}
	\begin{align}
		\eta_1^{\text{m}} &= \xi_1^{\text{m}} - \frac{1}{2}{\bf D}_{2'1}(\text{rfi}_2^{\text{m}}-\text{rfi}_{2'}^{\text{m}}), \\
		\eta_{1'}^{\text{m}} &= \xi_{1'}^{\text{m}} - \frac{1}{2}(\text{rfi}_{1'}^{\text{m}}-\text{rfi}_1^{\text{m}}).
	\end{align}
\end{subequations}
Similarly, the remaining expressions can be obtained by cyclic permutation of the indices. Now we can construct the intermediary variable $r_i$ to obtain the differential clock jitter, which is given by
\begin{equation}
	r_i = \frac{\eta_i^{\text{m}}}{k_j f_j^{\text{u}}}.
\end{equation}

Finally, we are able to compute the clock noise term with $r_i$ according to the TDI combination used for laser noise suppression, and subtract it from the TDI variable to remove the clock phase noise. The corresponding expressions for $X$ and $\zeta$ are presented in Appendix~\ref{appA}.

\section{\label{sec3}TTL Noise Model}
\subsection{TTL noise in TDI}

From Eqs.~(\ref{carrier_meas}) and (\ref{sideband_meas}), we know that TTL noise only exists in the ISI and TMI measurements. When we combine the ISI and TMI measurements, the OB displacement along the sensitive axis induced by the SC translational jitter, and part of the TTL contributions manifested as OB displacement caused by the angular jitters of SC and MOSAs, will be cancelled~\cite{Wanner2024}.  The SC and MOSA angular jitters can also cause OB jitter with respect to the TM for the reason that their pivots locate out of the OB. However, these TTL contributions appear simultaneously in ISI and TMI with opposite signs, which explains the cancellation in the single link readout $\xi_i$.

The TTL noise is often expressed in terms of length. Thus, we write the TTL noise in the three interferometers as
\begin{subequations}
	\label{n^TTL}
	\begin{align}
		n_{ji,\text{isi}}^{\text{TTL}} &= \frac{\nu_{j}}{c}(\rho_{i,\text{isi}}^{\text{RX}}+{\bf D}_{ji}\rho_{j,\text{isi}}^{\text{TX}}), \label{n^TTL-a} \\
		n_{i,\text{tmi}}^{\text{TTL}} &= \frac{\nu_{i}}{c}\rho_{i,\text{tmi}}, \label{n^TTL-b} \\
		n_{i,\text{rfi}}^{\text{TTL}} &= 0, \label{n^TTL-c}
	\end{align}
\end{subequations}
where $\rho_{i,\text{isi}}^{\text{RX}}$ and $\rho_{i,\text{isi}}^{\text{TX}}$ denote the TTL noise in the ISI received (RX) and transmitted (TX) beam, respectively; $\rho_{i,\text{tmi}}$ denotes the TTL noise in TMI. All $\rho$-terms are expressed in meters. Substituting Eq.~(\ref{n^TTL}) into Eq.~(\ref{carrier_meas}) and then Eq.~(\ref{xi}), we cancel the OB jitter due to the SC and MOSA motion, and obtain the final TTL contributions in the single link readout, given by
\begin{align}
	\xi_i^{\text{TTL}} &= \frac{\nu_{j}}{c}\left(\rho_{i,\text{isi}}^{\text{RX}}+\frac{1}{2}\rho_{i,\text{tmi}}+{\bf D}_{ji}\rho_{j,\text{isi}}^{\text{TX}}+\frac{1}{2}{\bf D}_{ji}\rho_{j,\text{tmi}}\right) \nonumber\\ &= \frac{\nu_{j}}{c}\left(\rho_i^{\text{RX}}+{\bf D}_{ji}\rho_j^{\text{TX}}\right),
\end{align}
with
\begin{subequations}
	\label{rho_def}
	\begin{align}
		\rho_i^{\text{RX}} &= \rho_{i,\text{isi}}^{\text{RX}}+\frac{1}{2}\rho_{i,\text{tmi}}, \\
		\rho_j^{\text{TX}} &= \rho_{j,\text{isi}}^{\text{TX}}+\frac{1}{2}\rho_{j,\text{tmi}}.
	\end{align}
\end{subequations}
According to Eqs.~(\ref{eta}) and (\ref{n^TTL-c}), we directly have $\eta_i^{\text{TTL}}=\xi_i^{\text{TTL}}$. To avoid confusion with the angular jitter in the following text, we will use $n_{ji}^{\text{TTL}}$ instead of $\eta_i^{\text{TTL}}$ to represent the TTL noise in $\eta_i$.

\begin{figure}[htbp]
\includegraphics[width=8.6cm]{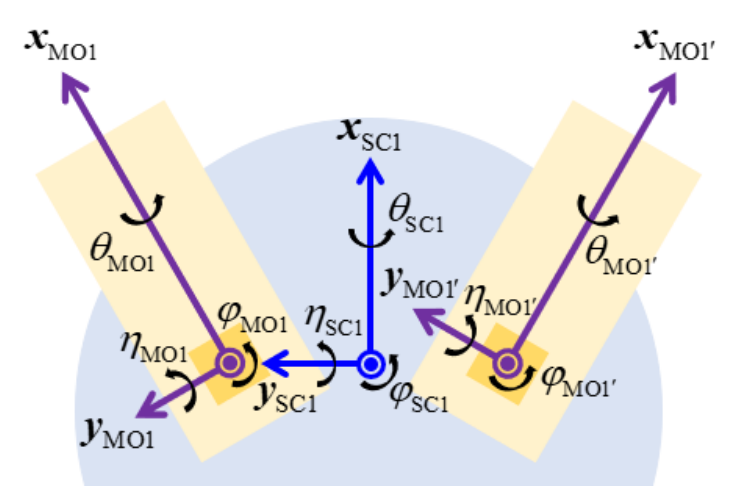}
\caption{\label{frames} Reference frames on SC1 as an example. The SC frame is shown in blue and the two MOSA frames are shown in purple. The opening angle between the two MOSAs, also known as the breathing angle, varies around $60^\circ$ due to the orbital mechanics of the triangular constellation.}
\end{figure}

The TTL noise is computed by the coupling coefficients and the angular jitters $(\tilde{\theta}_i,\tilde{\eta}_i,\tilde{\varphi}_i)$, also called roll, pitch and yaw angles, which can be computed as~\cite{George2023,Wanner2024}
\begin{subequations}
	\label{ang_jitter}
	\begin{align}
		\tilde{\theta}_i &= \tilde{\theta}_{\text{SC}i}\cos\gamma_i + \tilde{\eta}_{\text{SC}i}\sin\gamma_i + \tilde{\theta}_{\text{MO}i}, \\
		\tilde{\eta}_i &= \tilde{\eta}_{\text{SC}i}\cos\gamma_i - \tilde{\theta}_{\text{SC}i}\sin\gamma_i + \tilde{\eta}_{\text{MO}i}, \\
		\tilde{\varphi_i} &= \tilde{\varphi}_{\text{SC}i} + \tilde{\varphi}_{\text{MO}i},
	\end{align}
\end{subequations}
with $\text{SC}i'=\text{SC}i$. The angular jitters on the right-hand side are illustrated in Fig.~\ref{frames} where all angles are positive for counterclockwise rotation around the frame axes. The opening angle $\gamma_i$ varies around $30^\circ$ for the unprimed MOSA and around $-30^\circ$ for the primed MOSA. As for TTL noise, the roll angle $\tilde{\theta}_i$ is less important since any small rotation around the incoming beam axis will cause negligible TTL noise. Therefore, the TTL contributions $\rho_i^{\text{RX}}$ and $\rho_i^{\text{TX}}$ are expressed with the pitch and yaw angles, i.e.,
\begin{subequations}
	\begin{align}
		\rho_i^{\text{RX}} &= C_i^{\eta,\text{RX}}\tilde{\eta}_i + C_i^{\varphi,\text{RX}}\tilde{\varphi}_i, \\
		\rho_i^{\text{TX}} &= C_i^{\eta,\text{TX}}\tilde{\eta}_i + C_i^{\varphi,\text{TX}}\tilde{\varphi}_i,
	\end{align}
\end{subequations}
where the $C$-terms are the linear TTL coupling coefficients, satisfying
\begin{subequations}
	\begin{align}
		C_i^{\eta,\text{RX}} &= C_{i,\text{isi}}^{\eta,\text{RX}}+\frac{1}{2}C_{i,\text{tmi}}^{\eta}, \\
		C_i^{\varphi,\text{RX}} &= C_{i,\text{isi}}^{\varphi,\text{RX}}+\frac{1}{2}C_{i,\text{tmi}}^{\varphi}, \\
		C_i^{\eta,\text{TX}} &= C_{i,\text{isi}}^{\eta,\text{TX}}+\frac{1}{2}C_{i,\text{tmi}}^{\eta}. \\
		C_i^{\varphi,\text{TX}} &= C_{i,\text{isi}}^{\varphi,\text{TX}}+\frac{1}{2}C_{i,\text{tmi}}^{\varphi},
	\end{align}
\end{subequations}
corresponding to Eq.~(\ref{rho_def}).

Based on the discussion, the TTL noise in $\eta_i$ reads
\begin{align}
	n_{ji}^{\text{TTL}} &= \frac{\nu_j}{c}\left[C_i^{\eta,\text{RX}}\tilde{\eta}_i + C_i^{\varphi,\text{RX}}\tilde{\varphi}_i\right.  \nonumber\\ &\quad \left.+ {\bf D}_{ji}\left(C_i^{\eta,\text{TX}}\tilde{\eta}_i + C_i^{\varphi,\text{TX}}\tilde{\varphi}_i\right)\right]. \label{n_ji^TTL}
\end{align}
Propagating Eq.~(\ref{n_ji^TTL}) through a specific TDI combination yields the TTL contributions in the corresponding TDI variable. As an example, by replacing $\eta_i$ in Eq.~(\ref{X}) with $n_{ji}^{\text{TTL}}$, we can compute the TTL noise in $X$, written as $X_{\text{TTL}}$.

As for sideband postprocessing, we find the additional TTL noise is 7 orders of magnitude lower than that in the TDI variable by comparing the 5 -- 20 MHz beatnote frequency with the approximately 282 THz laser frequency. Therefore, the TTL noise introduced by the clock noise reduction algorithm is totally neglected.

\subsection{TTL noise estimation and subtraction}

We need to estimate the linear TTL coupling coefficients for TTL noise subtraction. In reality, TTL coefficients are expected to be slowly varying in flight. For this reason, we will consider linear drifts in TTL coefficients in the estimation process.

From Eq.~(\ref{n_ji^TTL}) we know that there will be 4 linear TTL coefficients to be estimated in a single link, and thus in total 24 for six links. If a linear TTL coefficient includes a linear drift, it can be written as
\begin{equation}
	C_i^{j,l}(t) = C_{i,0}^{j,l} + C_{i,1}^{j,l}\times(t-t_0)/86400 \label{coeff}
\end{equation}
where $C_{i,0}^{j,l}$ is the initial TTL coefficient at time $t_0$ and $C_{i,1}^{j,l}$ is the linear drift coefficient, with $j = \eta,\varphi$ and $l = \text{RX},\text{TX}$. In this case, the number of parameters to be estimated will be 48 for six links, including 24 TTL coefficients in mm/rad or $\upmu\text{m/rad}$ at time $t_0$ and 24 linear drift coefficients with the unit of $\upmu\text{m/rad/day}$.

However, we are not able to know the true angular jitters. Only the measured ones obtained by the DWS technique will be available which can be written as
\begin{subequations}
	\label{ang_meas}
	\begin{align}
		\hat{\tilde{\eta}}_i &= \tilde{\eta}_i + \delta\tilde{\eta}_i, \\
		\hat{\tilde{\varphi}}_i &= \tilde{\varphi}_i + \delta\tilde{\varphi}_i,
	\end{align}
\end{subequations}
where $\delta\tilde{\eta}_i$ and $\delta\tilde{\varphi}_i$ are the angular jitter sensing noise introduced in the ISI after performing calibration accounting for the magnification of the telescope and OB.

Here we use the TDI combination $X$ as an example to introduce the TTL noise estimation and subtraction procedure. To estimate the TTL coefficients at time $t_0$, we need to propagate each measured angular jitter through $X$ by replacing $\eta_i$ in Eq.~(\ref{X}) with $\hat{\tilde{\eta}}_i$ and $\hat{\tilde{\varphi}}_i$ in Eq.~(\ref{ang_meas}). The output is written as $\hat{X}_{i}^{j,l}$. As for linear drift coefficients, we introduce
\begin{subequations}
	\label{ang_t_meas}
	\begin{align}
		\hat{\tilde{\eta}}_{i,1} &= \hat{\tilde{\eta}}_i\times(t-t_0)/86400, \\
		\hat{\tilde{\varphi}}_{i,1} &= \hat{\tilde{\varphi}}_i\times(t-t_0)/86400.
	\end{align}
\end{subequations}
Similarly, we propagate them by replacing $\eta_i$ in Eq.~(\ref{X}) with $\hat{\tilde{\eta}}_{i,1}$ and $\hat{\tilde{\varphi}}_{i,1}$ in Eq.~(\ref{ang_t_meas}) and get the output $\hat{X}_{i,1}^{j,l}$. According to Eq.~(\ref{ang_meas}), we divide these TDI ouputs into true jitter terms and sensing noise terms:
\begin{subequations}
	\begin{align}
		\hat{X}_{i}^{j,l} &= X_{i}^{j,l} + \delta X_{i}^{j,l}, \\
		\hat{X}_{i,1}^{j,l} &= X_{i,1}^{j,l} + \delta X_{i,1}^{j,l}.
	\end{align}
\end{subequations}

We will perform least squares estimation for the TTL coupling parameters. The design matrix is composed of the TDI outputs $\hat{X}_{i}^{j,l}$ and $\hat{X}_{i,1}^{j,l}$. We use $\delta C_{i,0}^{j,l}$ to denote the estimation error of TTL coefficient $C_{i,0}^{j,l}$, and use $\delta C_{i,1}^{j,l}$ for the estimation error of linear drift coefficient $C_{i,1}^{j,l}$. With Eq.~(\ref{coeff}), the total error of a TTL coefficient at the time $t$ is computed as
\begin{equation}
	\delta C_i^{j,l}(t) = \delta C_{i,0}^{j,l} + \delta C_{i,1}^{j,l}\times(t-t_0)/86400. \label{delta_coeff}
\end{equation}
As a result, the estimated coefficient can be written as $\hat{C}_i^{j,l}(t) = C_i^{j,l}(t) + \delta C_i^{j,l}(t)$. Then the estimated TTL noise $\hat{X}^{\text{TTL}}$ reads
\begin{align}
	\hat{X}^{\text{TTL}} &= \sum_{i,j,l}\hat{C}_i^{j,l}\hat{X}_i^{j,l} \nonumber\\ &= \sum_{i,j,l}(C_i^{j,l} + \delta C_i^{j,l})(X_i^{j,l} + \delta X_i^{j,l}). \label{Xhat^TTL}
\end{align}

In this paper, TTL estimation is executed after clock noise suppression. We can subtract the estimated TTL noise from $X_{\text{corr}}^q$ derived in Appendix~\ref{appA}, i.e.,
\begin{equation}
	X_\text{corr}^{\text{TTL}} = X_\text{corr}^q - \hat{X}^{\text{TTL}}.
\end{equation}
$X_\text{corr}^{\text{TTL}}$ is the corrected TDI variable for $X$ as the laser noise, clock noise and TTL noise have been removed. Using Eq.~(\ref{Xhat^TTL}), the TTL estimation error in $X_\text{corr}^{\text{TTL}}$ can be computed as
\begin{align}
	\delta X^{\text{TTL}} &= X^{\text{TTL}} - \hat{X}^{\text{TTL}} \nonumber\\ &= \sum_{i,j,l}C_i^{j,l}X_i^{j,l} - \sum_{i,j,l}\hat{C}_i^{j,l}\hat{X}_i^{j,l} \nonumber\\ &= -\sum_{i,j,l}\delta C_i^{j,l}X_i^{j,l} - \sum_{i,j,l}\hat{C}_i^{j,l}\delta X_i^{j,l}. \label{deltaX_corr^TTL}
\end{align}
In Eq.~(\ref{deltaX_corr^TTL}) we find two contributions, one of which is caused by the coefficient estimation error while another is due to the sensing noise of the angular jitters.

Although we focus on $X$ in this section, the theoretical description is also valid for other TDI combinations including null channels. By replacing the notation $X$ with $\zeta$ in the above equations, one can obtain the analytical expressions for the null channel $\zeta$. In order to prevent GW signals from affecting TTL noise estimation, we suggest using the null channel to suppress laser phase noise and GW signals before estimating the TTL coefficients which will be used to remove the TTL noise in the intermediary variable $\eta_i$. Then we are able to construct other TDI variables for GW extraction, e.g., $X$, $Y$ and $Z$, that are free from TTL noise. This scheme will be used in the following simulation.

\section{\label{sec4}Simulation Setup}

In this section, we will introduce the setup of our numerical simulation which is performed to validate the effectiveness of the null channel $\zeta$ for the estimation of TTL coefficients with linear drifts. We focus on $\zeta$ exclusively because other null channels can be derived from $\zeta$ and can not give more information~\cite{Muratore2022,Muratore2023}.

We use TianQin's orbit introduced in~\cite{Ye2019} with a duration of 1 day. The sampling rates of all simulated data are 1 Hz. The order of Lagrange interpolation we use for TDI is 71. 

We use the amplitude spectral density (ASD) which is defined as the square root of power spectral density (PSD) to characterize the noise levels. The ASD of laser frequency noise is given by
\begin{equation}
	S_{\nu}^{1/2}(f) = 30\ \text{Hz/Hz}^{1/2}\times\text{NSF}(f)
\end{equation}
where the noise shape function reads
\begin{equation}
	\text{NSF}(f) = \sqrt{1+\left(\frac{4\ \text{mHz}}{f}\right)^4}.
\end{equation}
The ranging biases in TDI delays used for laser noise suppression are given in the order of 3 ns.

The fractional frequency noise ASDs of USO and sideband modulation are given by
\begin{equation}
	S_{\text{u}}^{1/2}(f) = 7.0\times10^{-14}\text{/Hz}^{1/2}\times\sqrt{\frac{1\ \text{Hz}}{f}}
\end{equation}
and
\begin{equation}
	S_{\text{m}}^{1/2}(f) = 2.7\times10^{-14}\text{/Hz}^{1/2}\times\text{NSF}(f).
\end{equation}

The contribution of polarization and stray light in each interferometer is characterized as equivalent displacement noise with an ASD of
\begin{equation}
	S_{\text{opt}}^{1/2}(f) = 0.2\ \text{pm/Hz}^{1/2}\times\text{NSF}(f).
\end{equation}

The readout noise levels in different carrier phase measurements are listed in terms of PSD as follows:
\begin{subequations}
	\begin{align}
		S_{\text{ro}}^{\text{isi}}(f) &= (0.50\ \text{pm/Hz}^{1/2}\times\text{NSF}(f))^2, \\
		S_{\text{ro}}^{\text{tmi}}(f) &= (0.15\ \text{pm/Hz}^{1/2}\times\text{NSF}(f))^2, \\
		S_{\text{ro}}^{\text{rfi}}(f) &= (0.15\ \text{pm/Hz}^{1/2}\times\text{NSF}(f))^2.
	\end{align}
\end{subequations}
Similarly, we have
\begin{subequations}
	\begin{align}
		S_{\text{ro,sb}}^{\text{isi}}(f) &= (7.15\ \text{pm/Hz}^{1/2}\times\text{NSF}(f))^2, \\
		S_{\text{ro,sb}}^{\text{tmi}}(f) &= (2.20\ \text{pm/Hz}^{1/2}\times\text{NSF}(f))^2, \\
		S_{\text{ro,sb}}^{\text{rfi}}(f) &= (2.20\ \text{pm/Hz}^{1/2}\times\text{NSF}(f))^2
	\end{align}
\end{subequations}
for sideband measurements.

The test mass acceleration noise is also included in our simulation and has an ASD of
\begin{equation}
	S_{\text{TM}}^{1/2}(f) = 1.0\ \text{fm/s}^2\text{/Hz}^{1/2}\times\sqrt{1+\left(\frac{0.1\ \text{mHz}}{f}\right)^2}.
\end{equation}

As for TTL noise, we assume that the ASDs of all SC angular jitters are
\begin{equation}
	S_{\text{SC}}^{1/2}(f) = 10\ \text{nrad/Hz}^{1/2}\times\text{NSF}(f). \label{ASD_ang_SC}
\end{equation}
We also model the MOSA jitter $\tilde{\eta}_i$ and $\tilde{\varphi}_i$ with respect to SC as follows:
\begin{subequations}
	\label{ASD_ang_MOSA}
	\begin{align}
		S_{\text{MO},\eta}^{1/2}(f) &= 1\ \text{nrad/Hz}^{1/2}\times\text{NSF}(f), \label{ASD_angy_MOSA}\\
		S_{\text{MO},\varphi}^{1/2}(f) &= 10\ \text{nrad/Hz}^{1/2}\times\text{NSF}(f). \label{ASD_angz_MOSA}
	\end{align}
\end{subequations}
Here the MOSA jitter $\tilde{\eta}_i$ is relatively smaller since only rotations around the $z$-axis are allowed for the MOSA pointing control. The angular jitter sensing noise is modeled as
\begin{equation}
	S_{\text{ang}}^{1/2}(f) = \frac{10}{M}\ \text{nrad/Hz}^{1/2}\times\text{NSF}(f)
\end{equation}
where $M = 300$ is the total magnification of the telescope and the imaging system on OB. The linear TTL coefficients at time $t_0$ are generated from a uniform distribution with a maximum absolute value of 3 mm/rad. The same distribution applies to the linear drift coefficients with a maximum absolute value of $100\ \upmu\text{m/rad/day}$.

\section{\label{sec5}Results}

Before we study TTL noise estimation, the algorithm for laser noise suppression and clock noise removal with the null channel $\zeta$ needs to be examined.

\begin{figure}[htbp]
\includegraphics[width=8.6cm]{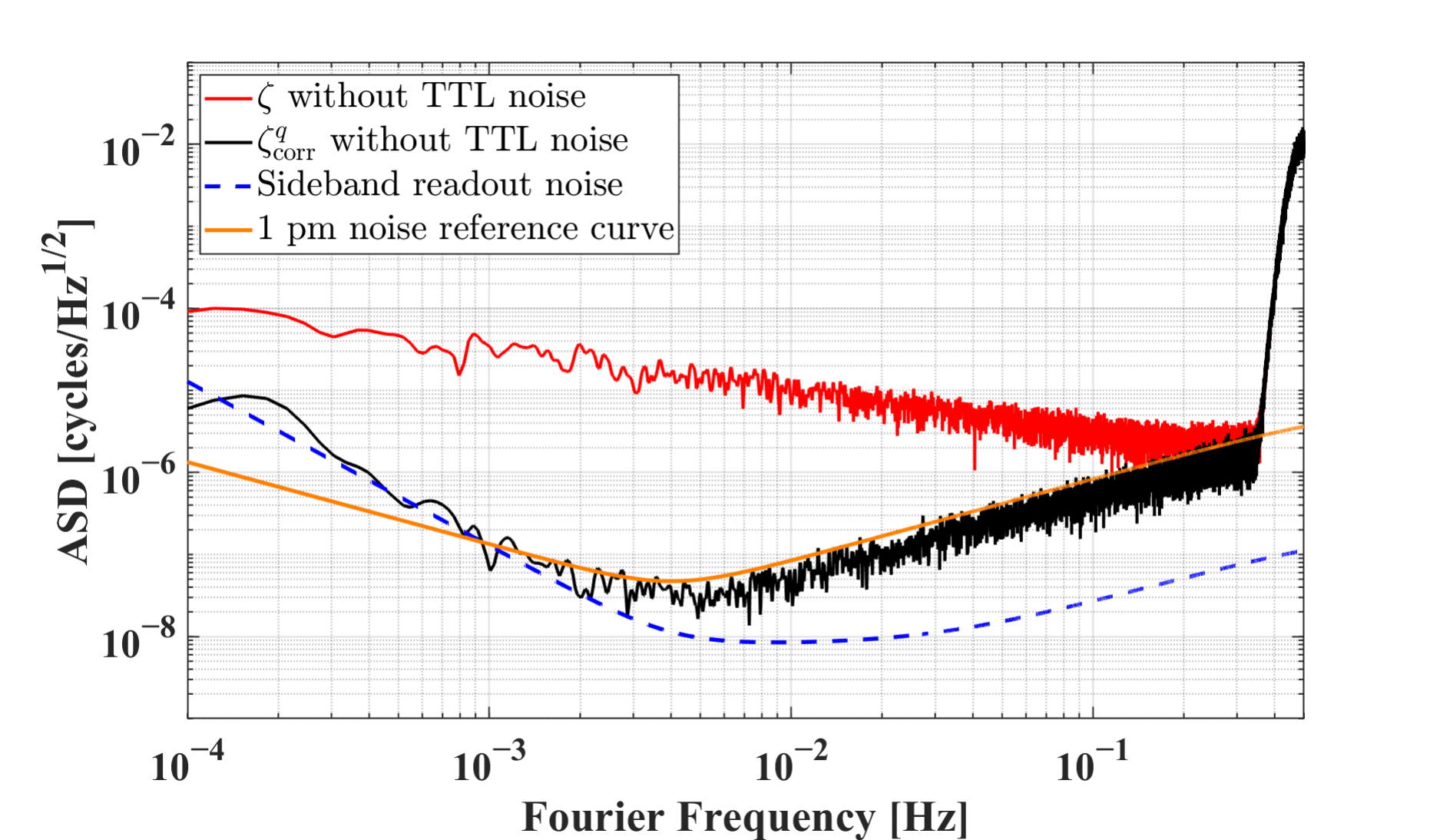}
\caption{\label{zeta_qc_noTTL} Residual noise levels after laser noise suppression (red) and clock noise reduction (black) with $\zeta$ in the absence of TTL noise. The green trace is higher than the 1 pm noise reference curve (orange) at frequencies below 1 mHz mainly due to the sideband readout noise (dashed blue).}
\end{figure}

\begin{figure}[htbp]
\includegraphics[width=8.6cm]{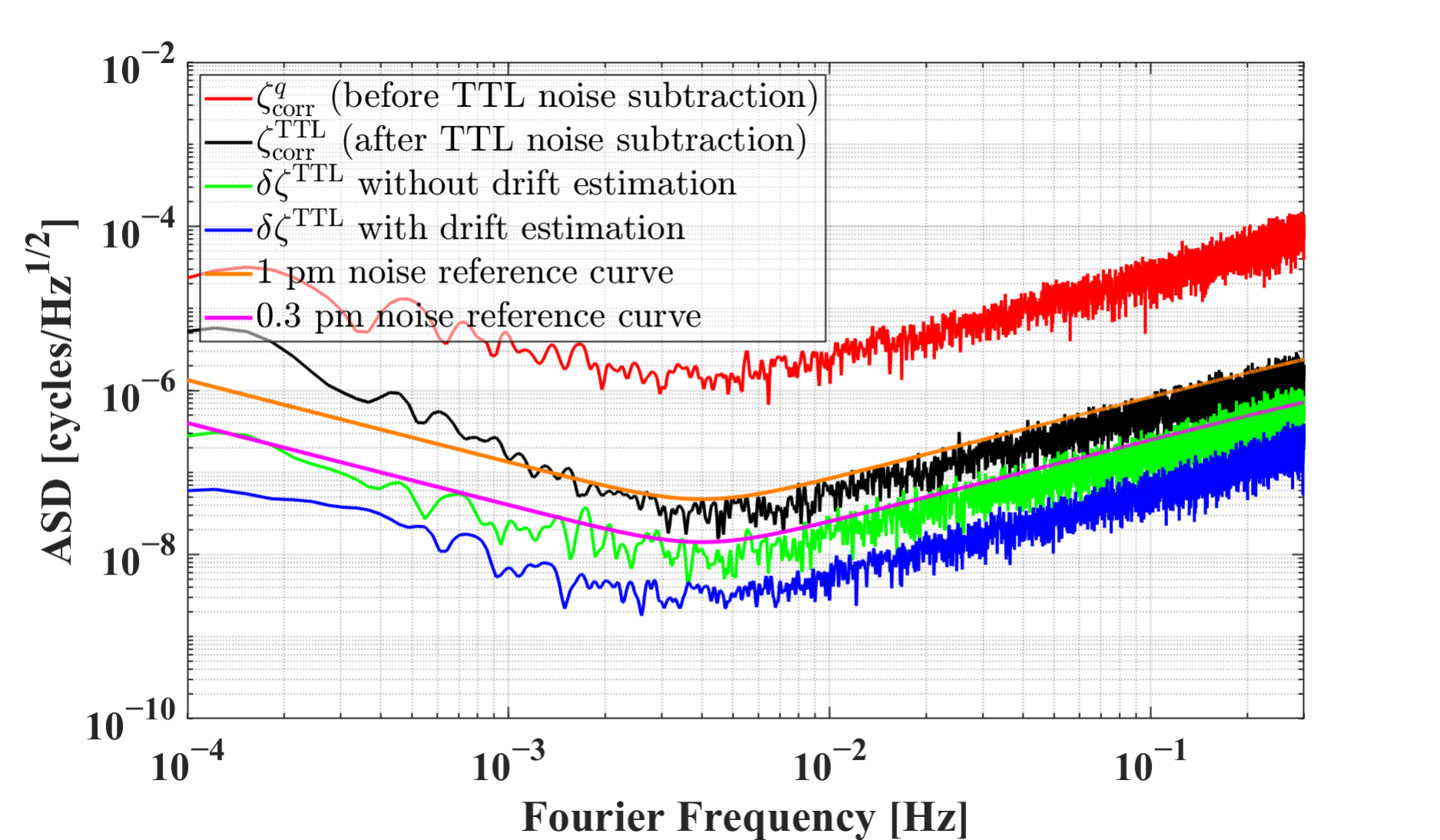}
\caption{\label{zeta_TTLc} TTL noise subtraction from $\zeta_\text{corr}^q$ (red). The residual noise and the TTL noise estimation error are shown in black and blue respectively. The green trace represents the estimation error in the case that drift coefficients are not included in TTL noise estimation and subtraction.}
\end{figure}

Fig.~\ref{zeta_qc_noTTL} shows the simulation results with all noise enabled except TTL noise. The blue trace represents the noise level in $\zeta$. After removing clock noise, we expect the residual noise level (black) will be within the 1 pm noise allocation (orange) whose PSD is given by
\begin{equation}
	S_{1\,\text{pm}}^{\zeta}(f) = \frac{24\sin^2(\pi f d_0)}{\lambda^2}\times1\ \text{pm}^2\text{/Hz}\times(\text{NSF}(f))^2 \label{PSD_1pm_zeta}
\end{equation}
where $\lambda = 1064.5\ \text{nm}$ is the laser frequency and $d_0 = \sqrt{3}\times 10^8\ \text{m}/c$ is the nominal long-arm delay for TianQin. The 0.3 pm noise reference curve used in Fig.~\ref{zeta_TTLc} can be computed by replacing $1\ \text{pm/Hz}^{1/2}$ in Eq.~(\ref{PSD_1pm_zeta}) with $0.3\ \text{pm/Hz}^{1/2}$. From Fig.~\ref{zeta_qc_noTTL} we can see the significant interpolation error at frequencies above 0.3 Hz, and find the sideband readout noise (dashed blue) will be dominant at frequencies below 1 mHz based on Eq.~(\ref{PSD_zeta_ro_sb}). This will not be a serious problem since the sampling rate after decimation will be higher than 1 Hz in practice, and we expect that the actual slope of readout noise at low frequencies will be much better than $f^{-2}$.

\begin{table}[htbp]
	\caption{\label{drift_coeff}Linear drift coefficient values and errors.}
	\begin{ruledtabular}
		\begin{tabular}{cD{.}{.}{5}D{.}{.}{6}D{.}{.}{6}}
			Coefficient & \multicolumn{1}{c}{True Value} & \multicolumn{1}{c}{Estimated Value} & \multicolumn{1}{c}{Error}\\
			 & \multicolumn{1}{c}{$[\upmu\text{m/rad/day}]$} & \multicolumn{1}{c}{$[\upmu\text{m/rad/day}]$} & \multicolumn{1}{c}{$[\upmu\text{m/rad/day}]$}\\
			\colrule
			$C_{1,1}^{\eta,\text{RX}}$ 	&  27.6 &  32.5 &  4.9\\
			$C_{1',1}^{\eta,\text{RX}}$ 	&  16.5 &  19.1 &  2.6\\
			$C_{2,1}^{\eta,\text{RX}}$ 	& -40.3 & -34.7 &  5.5\\
			$C_{2',1}^{\eta,\text{RX}}$ 	&  -5.4 &  -2.4 &  3.1\\
			$C_{3,1}^{\eta,\text{RX}}$ 	& -23.9 & -16.0 &  7.9\\
			$C_{3',1}^{\eta,\text{RX}}$ 	& -45.7 & -43.9 &  1.8\\
			$C_{1,1}^{\varphi,\text{RX}}$	&  51.3 &  49.9 & -1.4\\
			$C_{1',1}^{\varphi,\text{RX}}$	&  94.7 &  93.1 & -1.5\\
			$C_{2,1}^{\varphi,\text{RX}}$ 	&  32.4 &  29.5 & -2.9\\
			$C_{2',1}^{\varphi,\text{RX}}$	&  78.5 &  76.4 & -2.1\\
			$C_{3,1}^{\varphi,\text{RX}}$	& -68.0 & -70.1 & -2.0\\
			$C_{3',1}^{\varphi,\text{RX}}$	& -38.4 & -40.7 & -2.3\\
			$C_{1,1}^{\eta,\text{TX}}$ 	& -91.4 & -86.9 &  4.5\\
			$C_{1',1}^{\eta,\text{TX}}$ 	& -24.5 & -21.0 &  3.5\\
			$C_{2,1}^{\eta,\text{TX}}$ 	& -93.9 & -93.2 &  0.7\\
			$C_{2',1}^{\eta,\text{TX}}$ 	&  19.6 &  22.2 &  2.6\\
			$C_{3,1}^{\eta,\text{TX}}$ 	&  36.8 &  44.7 &  7.9\\
			$C_{3',1}^{\eta,\text{TX}}$ 	& -14.4 & -11.0 &  3.5\\
			$C_{1,1}^{\varphi,\text{TX}}$ 	&  68.7 &  67.5 & -1.2\\
			$C_{1',1}^{\varphi,\text{TX}}$ 	&  -1.4 &  -3.9 & -2.5\\
			$C_{2,1}^{\varphi,\text{TX}}$ 	& -54.1 & -57.1 & -3.0\\
			$C_{2',1}^{\varphi,\text{TX}}$ 	&  39.6 &  38.4 & -1.1\\
			$C_{3,1}^{\varphi,\text{TX}}$ 	& -41.0 & -44.6 & -3.6\\
			$C_{3',1}^{\varphi,\text{TX}}$ 	&  -9.7 & -12.4 & -2.7\\
		\end{tabular}
	\end{ruledtabular}
\end{table}

A band-pass filter with cutoff frequencies of 5 mHz and 0.3 Hz is applied to the TDI outputs of phase and angular jitters in TTL coefficient estimation to remove disturbances from the interpolation error and sideband readout noise. The estimated linear drift coefficient values and errors are listed in Table~\ref{drift_coeff}. The error is defined as the difference between the estimated and true value. The accuracy of all 24 estimated linear TTL coefficients is within $10\ \upmu\text{m/rad}$, and the errors of all 24 linear drift coefficients listed in Table~\ref{drift_coeff} are below $10\ \upmu\text{m/rad/day}$. With these estimated parameters, TTL noise can be removed from $\zeta_{\text{corr}}^q$ shown in red in Fig.~\ref{zeta_TTLc}, and the output $\zeta_{\text{corr}}^{\text{TTL}}$ is shown in black. The blue trace represents the TTL noise estimation error. Here we discuss another scenario that drift coefficients are not estimated for TTL subtraction and the corresponding TTL estimation error is shown in green. We find the green trace approaches the 0.3 pm noise referenece curve (magenta) while the blue trace is significantly lower.

\begin{figure}[htbp]
\includegraphics[width=8.6cm]{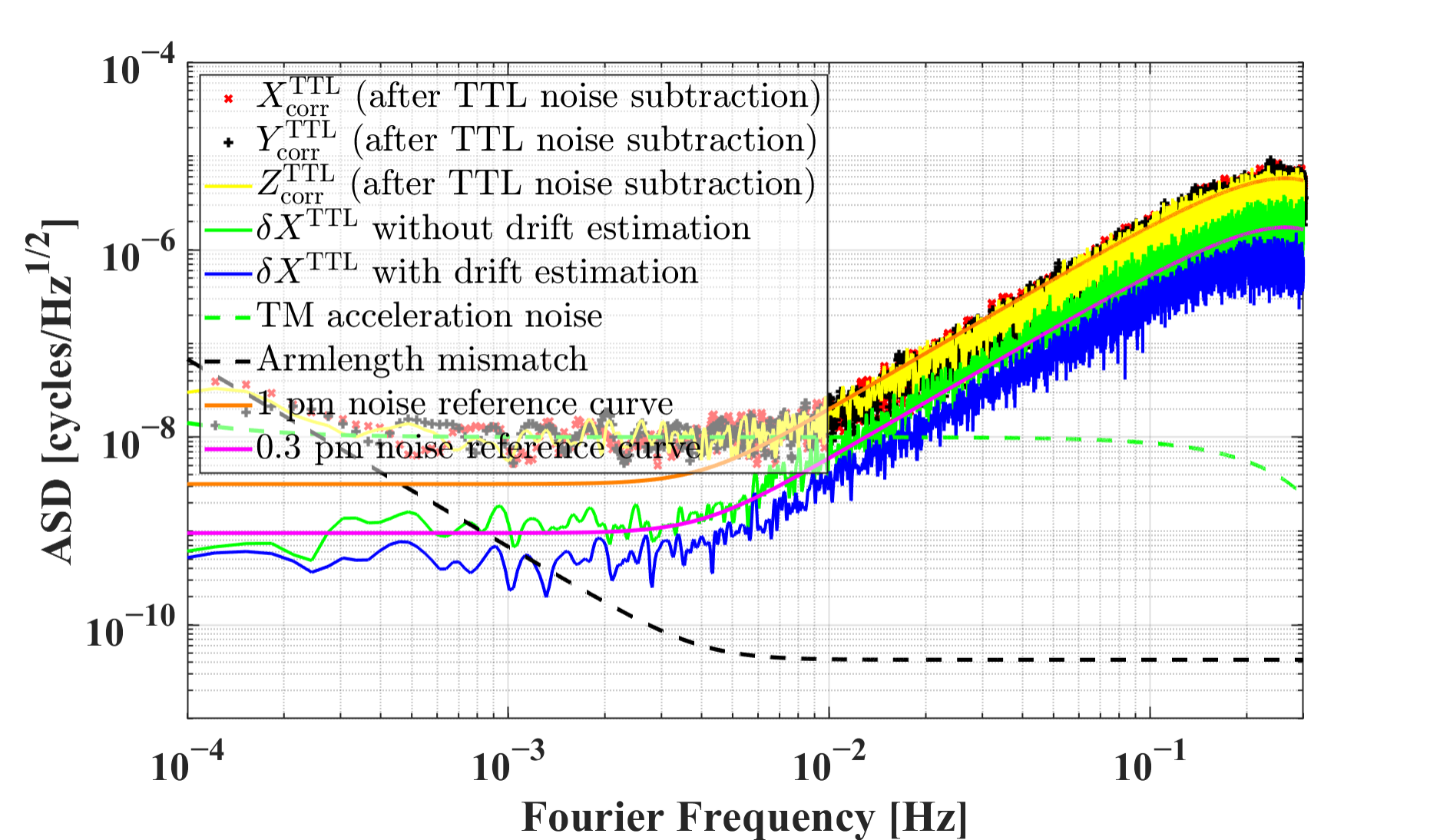}
\caption{\label{X_TTLc} Residual noise level in $X$, $Y$ and $Z$ after subtracting the TTL noise and clock noise. Below 7 mHz it is dominated by TM acceleration noise (dashed green) and armlength mismatch of the virtual interferometer for TDI (dashed black). The blue and solid green traces represent TTL noise residuals in $X$ with and without drift estimation.}
\end{figure}

Since the Michelson combinations (or the equivalent orthogonal combinations derived by them) are commonly used for GW extraction, we need to subtract the TTL noise from the intermediary variable $\eta_i$ with the estimated coefficients, and then propagate $\eta_i$ through $X$. The same can be applied to $Y$ and $Z$ combinations by cyclic permutation of the indices. After clock noise removal, we obtain the TDI variables in Fig.~\ref{X_TTLc} which are not limited by the TTL noise, but are higher than the 1 pm noise reference curve for $X$ (orange) at frequencies below 7 mHz as expected mainly due to the TM acceleration noise (dashed green) as well as the armlength mismatch of the constructed virtual interferometer for TDI (dashed black). The PSD model of TM acceleration noise in $X$ is given by Eq.~(\ref{PSD_X_TM}) in Appendix~\ref{appA}. The armlength (or delay) mismatch of $X$ can be computed by
\begin{align}
	\Delta d &\approx 2d_0[(\dot{d}_{2'1}+\dot{d}_{12'})^2-(\dot{d}_{31'}+\dot{d}_{1'3})^2] \nonumber\\ &\quad + 8d_0^2(\ddot{d}_{2'1}+\ddot{d}_{12'}-\ddot{d}_{31'}-\ddot{d}_{1'3}).
\end{align}
The dashed black trace is plotted with the following formula:
\begin{equation}
	S_{\Delta d}^{1/2}(f) = \sqrt{4\sin^2(\pi f \Delta d)} S_{\nu}^{1/2}(f).
\end{equation}
The 1 pm noise reference curve for $X$ is generated by
\begin{equation}
	S_{1\,\text{pm}}^{X}(f) = \frac{4C_{XX}}{\lambda^2}\times1\ \text{pm}^2\text{/Hz}\times(\text{NSF}(f))^2 \label{PSD_1pm_X}
\end{equation}
where
\begin{equation}
	C_{XX} = 16\sin^2(4\pi f d_0)\sin^2(2\pi f d_0).
\end{equation}
By replacing $1\ \text{pm/Hz}^{1/2}$ in Eq.~(\ref{PSD_1pm_X}) with $0.3\ \text{pm/Hz}^{1/2}$, we obtain the 0.3 pm noise reference curve for $X$ (magenta). We find the green trace exceeds the 0.3 pm noise reference curve while the blue trace is below. Therefore, linear drift coefficients may need to be considered for a better TTL noise correction.

Based on the results, we believe the TTL noise can be sufficiently reduced using the coefficients estimated in the null channel $\zeta$. With a better laser frequency stablization, the TDI output will meet TianQin's requirement and can be used for GW data analysis.

\section{\label{sec6}Discussion}

The exact value of each linear drift coefficient is not yet clear since it may be influenced by various factors such as pointing control and temperature change. We will discuss the influence of pointing error in Sec.~\ref{sec6A}.

Both SC and MOSA angular jitters contribute to the TTL noise. In general, we assume all MOSA angular jitters are independent. However, MOSA$i$ and MOSA$i'$ share the same angular jitter of SC$i$. Specifically, the SC yaw jitter contributes equally to the two onboard MOSAs according to Eq.~(\ref{ang_jitter}) because of the parallel $z$-axes of the SC and MOSAs. This will lead to a strong correlation between coefficients associated with the same SC yaw jitter, which will degrade the accuracy of the estimated TTL coefficients. Therefore, the levels of the uncorrelated MOSA yaw jitters will significantly affect the TTL estimation results. This topic has been discussed for $X$ combination in~\cite{Paczkowski2022,George2023}, and we will show the corresponding results for the null channel $\zeta$ in Sec.~\ref{sec6B}.

\subsection{\label{sec6A}Pointing error}

\begin{figure}[htbp]
\includegraphics[width=8.6cm]{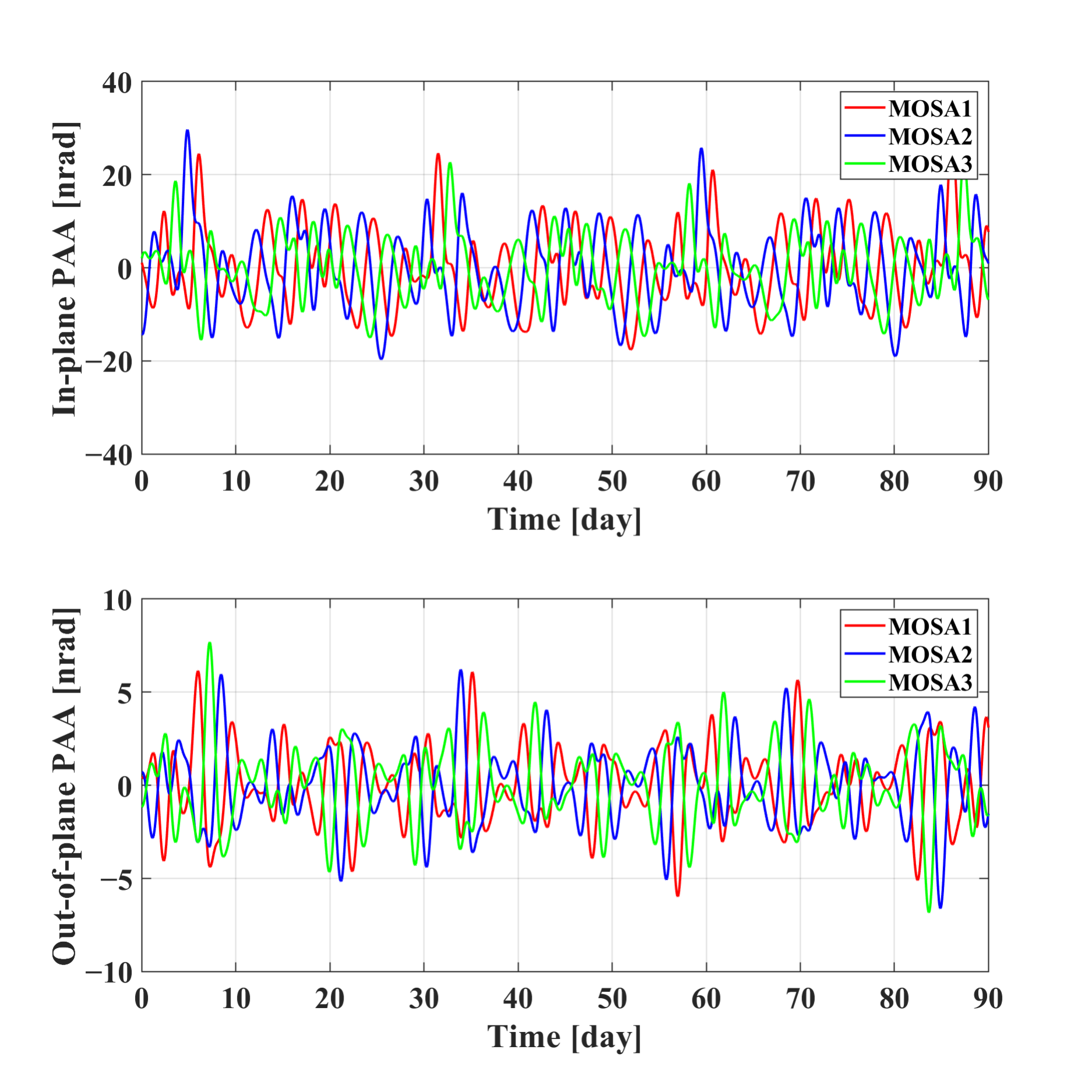}
\caption{\label{PAA_in_out} Unprimed in-plane (upper plot) and out-of-plane PAAs (lower plot) of 90 days in TianQin. The mean value of each in-plane PAA (approximately $-23.07\ \upmu\text{rad}$) has been subtracted.}
\end{figure}

Due to the light travel time and the relative lateral motion among the three SC, the received beam and transmitted beam of each MOSA do not coincide. The angle between the received beam and transmitted beam is referred to as point-ahead angle (PAA). Generally, the PAAs in the primed MOSAs, i.e., PAA$2'$, PAA$3'$ and PAA$1'$, approximate to the unprimed PAAs, i.e., PAA1, PAA2 and PAA3, respectively. The local constellation plane of each SC is defined by the received beams from the two remote SC. Therefore, all PAAs can be decomposed into in-plane PAAs and out-of-plane PAAs. The unprimed PAAs of a three-month period in TianQin are shown in Fig.~\ref{PAA_in_out}.

The out-of-plane PAAs in TianQin are within the range of $\pm 10\ \text{nrad}$ and thus no dedicated compensation is needed. However, since the in-plane PAAs are approximately $23.07\ \upmu\text{rad}$ (or $-23.07\ \upmu\text{rad}$), compensation for the in-plane PAAs in TianQin is required. Considering that the in-plane PAA variations are only in the order of $10\ \text{nrad}$, a static compensation is sufficient in TianQin~\cite{Wang2024}. As a result, the in-plane and out-of-plane PAA variations are manifested as slowly varying pointing errors, causing drifts in TTL coefficients.

If a TX beam has a wavefront error (WFE), the beam pointing jitter will cause TTL noise. According to~\cite{Bender2005}, an approximate expression of the TTL noise caused by wavefront tilt angles $\alpha$ and $\beta$ reads
\begin{equation}
	s \approx \frac{1}{E_{\text{c}}}(C_1\alpha+C_2\beta+C_3\alpha^2+C_4\beta^2+C_5\alpha\beta) \label{TTL_p}
\end{equation}
with
\begin{equation}
	E_{\text{c}} = 1 - \frac{2\pi^2}{\lambda^2}w_{\text{rms}}^2,
\end{equation}
where $w_{\text{rms}}$ is the root mean square (RMS) of the WFE. $C_1/E_{\text{c}}$ and $C_2/E_{\text{c}}$ are linear TTL coefficients that have been considered in the above simulation, while the three quadratic TTL coefficients $C_3/E_{\text{c}}$, $C_4/E_{\text{c}}$ and $C_5/E_{\text{c}}$ will be discussed in the following. We divide the wavefront tilt angles $\alpha_i$ and $\beta_i$ of MOSA$i$ into two components,
\begin{subequations}
	\label{alpha,beta}
	\begin{align}
		\alpha_i &= \text{PAA}_i^{\text{in}} + \tilde{\varphi}_i, \\
		\beta_i &= \text{PAA}_i^{\text{out}} + \tilde{\eta}_i,
	\end{align}
\end{subequations}
where $\text{PAA}_i^{\text{in}}$ and $\text{PAA}_i^{\text{out}}$ denote the in-plane and out-of-plane PAAs respectively. Substituting Eq.~(\ref{alpha,beta}) into the quadratic terms in Eq.~(\ref{TTL_p}) and neglecting the out-of-band components, we have
\begin{equation}
	s_2^{\epsilon} \approx \frac{1}{E_{\text{c}}}\left[\begin{array}{l}C_3\tilde{\varphi}_i^2 + C_4\tilde{\eta}_i^2 + C_5\tilde{\varphi}_i\tilde{\eta}_i \\ + (2C_3\text{PAA}_i^{\text{in}}+C_5\text{PAA}_i^{\text{out}})\tilde{\varphi}_i \\ + (2C_4\text{PAA}_i^{\text{out}}+C_5\text{PAA}_i^{\text{in}})\tilde{\eta}_i\end{array}\right], \label{s_2^epsilon}
\end{equation}
where terms in the first row in the square bracket are quadratic, and the remaining terms are linear terms with coefficient drifts caused by PAA variations.

WFE can be expressed as a linear combination of Zernike polynomials. If one considers the eight Zernike polynomials given in~\cite{Bender2005}, we have
\begin{subequations}
	\label{C345}
	\begin{align}
		C_3 &= \frac{2\pi^2}{\sqrt{3}\lambda^2}a_{\text{TX}}^2\left(\frac{1}{4}b_2^0 + \frac{1}{\sqrt{8}}b_2^2\right), \\
		C_4 &= \frac{2\pi^2}{\sqrt{3}\lambda^2}a_{\text{TX}}^2\left(\frac{1}{4}b_2^0 - \frac{1}{\sqrt{8}}b_2^2\right), \\
		C_5 &= \frac{2\pi^2}{\sqrt{3}\lambda^2}a_{\text{TX}}^2\left(\frac{1}{\sqrt{2}}b_2^{-2}\right),
	\end{align}
\end{subequations}
where $a_{\text{TX}}$ is the radius of TX aperture; $b_2^0$ and $b_2^{\pm2}$ are amplitudes of certain Zernike polynomials corresponding to defocus and primary astigmatism. One can find how the amplitudes of certain aberrations contribute to the TTL noise by substituting Eq.~(\ref{C345}) into Eq.~(\ref{s_2^epsilon}).

In the following simuations, we assume $w_{\text{rms}}=0.05\lambda$, $a_{\text{TX}}=0.15\ \text{m}$, $b_2^0=0.025\lambda$ and $b_2^{\pm2}=0.0125\lambda$. We remove the NSFs in Eq.~(\ref{ASD_ang_SC}) and (\ref{ASD_ang_MOSA}) and consider white angular jitters. The initial linear TTL coefficients are still generated from a uniform distribution with a maximum absolute value of 3 mm/rad. TTL noise introduced by Eq.~(\ref{s_2^epsilon}) is also included in the simulations, but linear drifts are not added. In addition to the linear TTL coefficients and linear drift coefficients, the quadratic TTL coefficients are also considered in TTL noise estimation.

\begin{figure}[htbp]
\includegraphics[width=8.6cm]{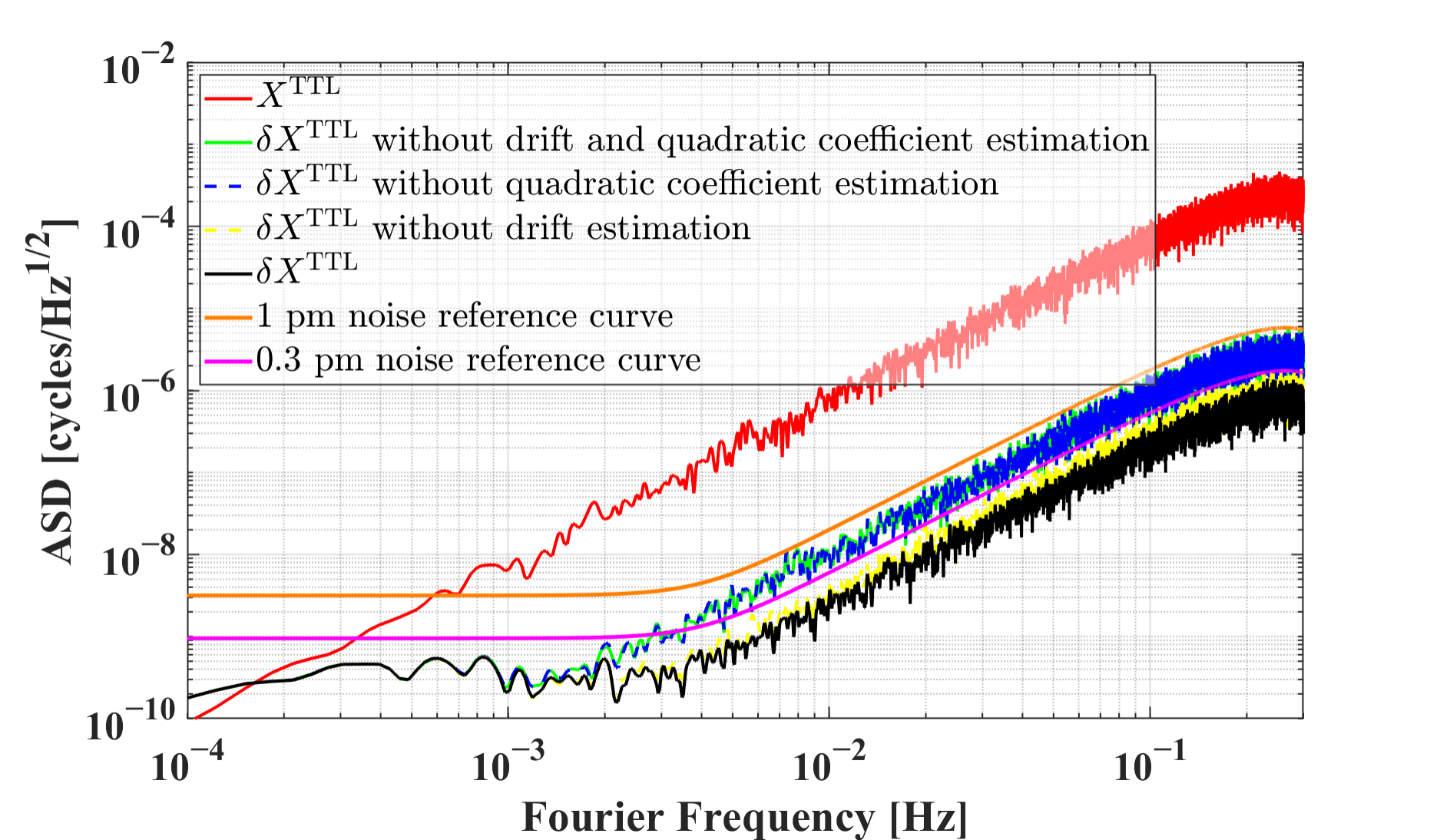}
\caption{\label{zeta_TTLc_p_day1} TTL noise in $X$ with PAAs on day 1, and TTL noise residuals under different conditions after subtraction.}
\end{figure}

\begin{figure}[htbp]
\includegraphics[width=8.6cm]{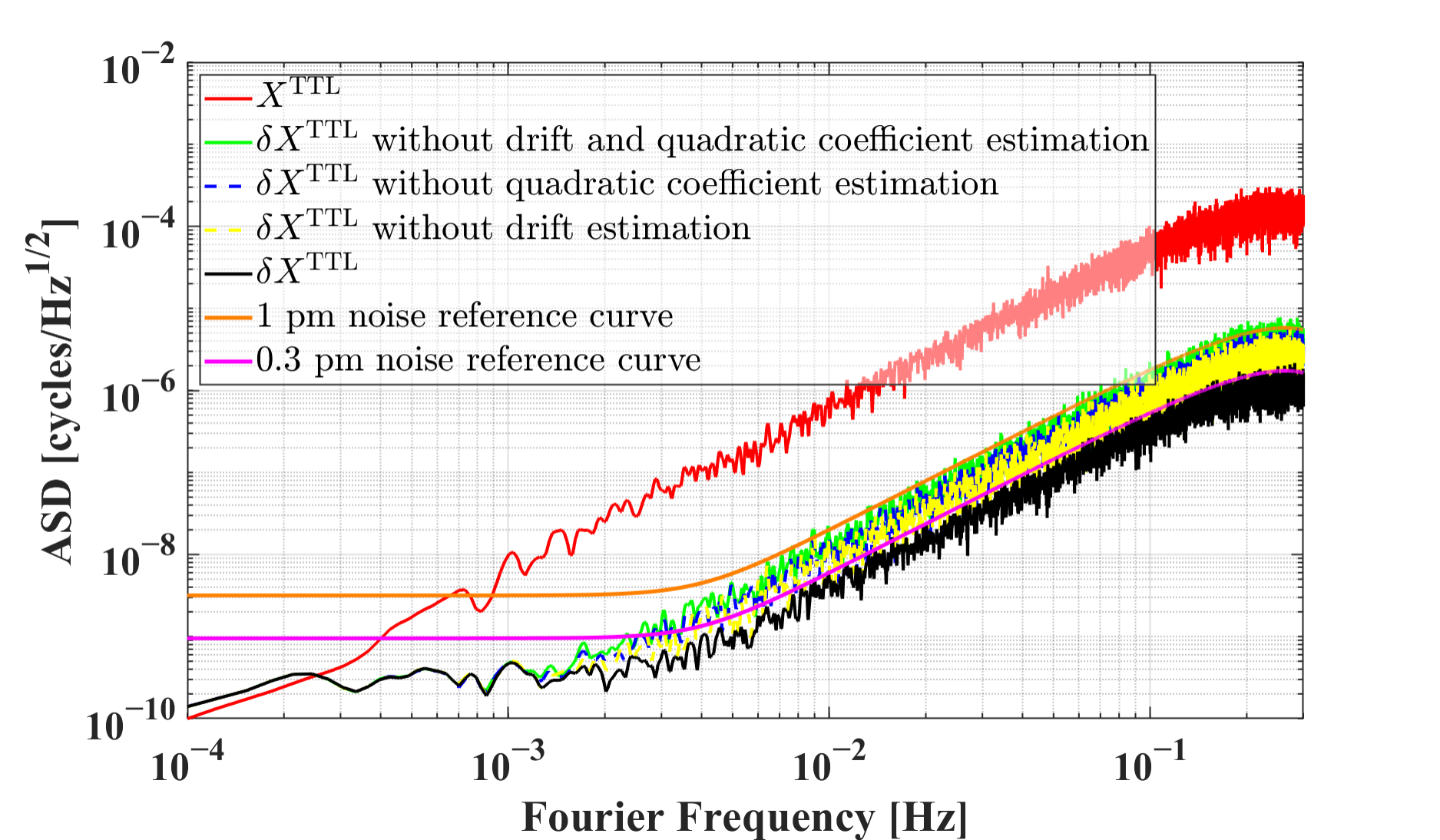}
\caption{\label{zeta_TTLc_p_day32} TTL noise in $X$ with PAAs on day 32, and TTL noise residuals under different conditions after subtraction.}
\end{figure}

Fig.~\ref{zeta_TTLc_p_day1} shows the TTL noise subtraction results in $X$ with the PAAs on day 1. The total TTL noise is shown in red. If one considers only linear TTL coefficients for TTL noise estimation and subtraction, the TTL noise residual (green) is slightly below the 1 pm noise referece curve at high frequencies. Comparing the dashed blue and yellow traces, we find the contribution of quadratic coefficients is much higher than the one of coefficient drifts caused by PAA variations. If all linear and quadratic TTL coefficients as well as linear drift coefficients are estimated, we find the TTL noise residual (black) is below the 0.3 pm noise reference curve (magenta).

We find from Fig.~\ref{PAA_in_out} that in-plane PAAs can vary violently at some time. Taking day 32 as an example, we use PAAs on this day for another simulation with the same setup described in this section. The results are shown in Fig.~\ref{zeta_TTLc_p_day32}. We find the TTL noise levels caused by coefficient drifts (yellow) and quadratic coefficients (blue) are similar. The yellow and black traces are higher than the ones in Fig.~\ref{zeta_TTLc_p_day1} due to the violent and nonlinear PAA variations. Nonetheless, the best results of TTL noise subtraction (black) in both scenarios meet the requirement (magenta). This implies that both linear drift coefficients and quadratic TTL coefficients might need to be taken into account for TTL noise subtraction.

\subsection{\label{sec6B}MOSA yaw jitter}

In this section, We will investigate the influence of MOSA yaw jitter in the null channel $\zeta$ by four independent simulations among which the MOSA yaw jitter level is set to 0, 1, 3, 10 $\text{nrad/Hz}^{1/2}$ respectively. The SC angular jitters as well as MOSA pitch jitter are still modeled as Eqs.~(\ref{ASD_ang_SC}) and (\ref{ASD_angy_MOSA}) respectively. Every input linear TTL coefficient at time $t_0$ is exactly 3 mm/rad and every linear drift coefficient is exactly $100\ \upmu\text{m/rad/day}$. TTL noise introduced by Eq.~(\ref{s_2^epsilon}) is not considered.

\begin{table}[htbp]
	\caption{\label{coeff_error_angz_MOSA}Estimation errors of TTL coefficients at time $t_0$. The MOSA yaw jitter levels in cases 1 to 4 are 0, 1, 3, 10 $\text{nrad/Hz}^{1/2}$ respectively.}
	\begin{ruledtabular}
		\begin{tabular}{cD{.}{.}{2}D{.}{.}{2}D{.}{.}{2}D{.}{.}{2}}
			Error & \multicolumn{4}{c}{Value [$\upmu\text{m/rad}$]}\\
			\cline{2-5}
			 & \multicolumn{1}{c}{Case 1} & \multicolumn{1}{c}{Case 2} & \multicolumn{1}{c}{Case 3} & \multicolumn{1}{c}{Case 4}\\
			\colrule
			$\delta C_{1,0}^{\eta,\text{RX}}$		&    -4.1 &  -0.5 &   4.2 & -3.6\\
			$\delta C_{1',0}^{\eta,\text{RX}}$	 	&    -0.5 &  -0.2 &   0.8 & -8.7\\
			$\delta C_{2,0}^{\eta,\text{RX}}$ 		&    -2.6 &  -1.2 &   2.5 & -0.6\\
			$\delta C_{2',0}^{\eta,\text{RX}}$		&     0.4 &   1.2 &  -0.0 &  1.6\\
			$\delta C_{3,0}^{\eta,\text{RX}}$ 		&     2.2 &   4.6 &  -0.7 &  1.8\\
			$\delta C_{3',0}^{\eta,\text{RX}}$ 		&     2.9 &   5.8 &  -2.7 & -2.4\\
			$\delta C_{1,0}^{\varphi,\text{RX}}$  	& -4120.1 &  -5.6 &   0.8 & -0.1\\
			$\delta C_{1',0}^{\varphi,\text{RX}}$ 	& -4119.3 &  -3.1 &  -0.1 &  1.5\\
			$\delta C_{2,0}^{\varphi,\text{RX}}$  	& -2572.6 & -10.5 &   7.8 &  3.0\\
			$\delta C_{2',0}^{\varphi,\text{RX}}$ 	& -2574.4 & -11.8 &  11.2 &  3.7\\
			$\delta C_{3,0}^{\varphi,\text{RX}}$  	& -2561.4 &  22.2 &  -5.8 &  2.2\\
			$\delta C_{3',0}^{\varphi,\text{RX}}$ 	& -2559.5 &  23.5 &  -4.5 & -0.1\\
			$\delta C_{1,0}^{\eta,\text{TX}}$ 		&    -2.5 &  -1.3 &   2.7 &  2.6\\
			$\delta C_{1',0}^{\eta,\text{TX}}$ 		&     0.4 &  -0.9 &   0.2 & -1.0\\
			$\delta C_{2,0}^{\eta,\text{TX}}$ 		&    -4.5 &  -0.4 &   2.1 & -1.5\\
			$\delta C_{2',0}^{\eta,\text{TX}}$ 		&    -3.1 &  -0.0 &  -2.8 &  1.1\\
			$\delta C_{3,0}^{\eta,\text{TX}}$ 		&     1.9 &   2.7 &   1.7 &  1.1\\
			$\delta C_{3',0}^{\eta,\text{TX}}$ 		&     1.0 &   6.4 &  -0.7 & -3.3\\
			$\delta C_{1,0}^{\varphi,\text{TX}}$ 		& -3558.1 &   8.7 &   0.0 &  1.3\\
			$\delta C_{1',0}^{\varphi,\text{TX}}$ 	& -3556.3 &  12.5 &  -2.5 &  1.7\\
			$\delta C_{2,0}^{\varphi,\text{TX}}$ 		& -2350.8 &  17.3 &   4.1 &  1.0\\
			$\delta C_{2',0}^{\varphi,\text{TX}}$ 	& -2351.2 &  14.4 &   6.8 &  5.6\\
			$\delta C_{3,0}^{\varphi,\text{TX}}$ 		& -2875.0 &  11.2 & -12.1 &  2.0\\
			$\delta C_{3',0}^{\varphi,\text{TX}}$ 	& -2874.3 &  12.1 & -10.6 & -1.6\\
			RMS 								&  2171.7 &  10.1 &   5.1 &  2.9\\
		\end{tabular}
	\end{ruledtabular}
\end{table}

The errors of the estimated TTL coefficients at time $t_0$ are listed in Table~\ref{coeff_error_angz_MOSA}. We find unacceptably large errors in the coefficients related to $\tilde{\varphi}$ in the absence of the independent MOSA yaw jitters due to the strong correlation between $\tilde{\varphi}_i$ and $\tilde{\varphi}_{i'}$ on SC$i$. For a better comparison, we compute the RMS values of $\delta C_{i,0}^{j,l}$ for each case with the following formula:
\begin{equation}
	\text{RMS} = \sqrt{\frac{1}{24}\sum_{i,j,l}(\delta C_{i,0}^{j,l})^2}.
\end{equation}
As the MOSA yaw jitter level increases, the RMS of coefficient error shown in Table~\ref{coeff_error_angz_MOSA} will decrease, which is in agreement with the discussion in~\cite{Paczkowski2022}.

\begin{figure}[htbp]
\includegraphics[width=8.6cm]{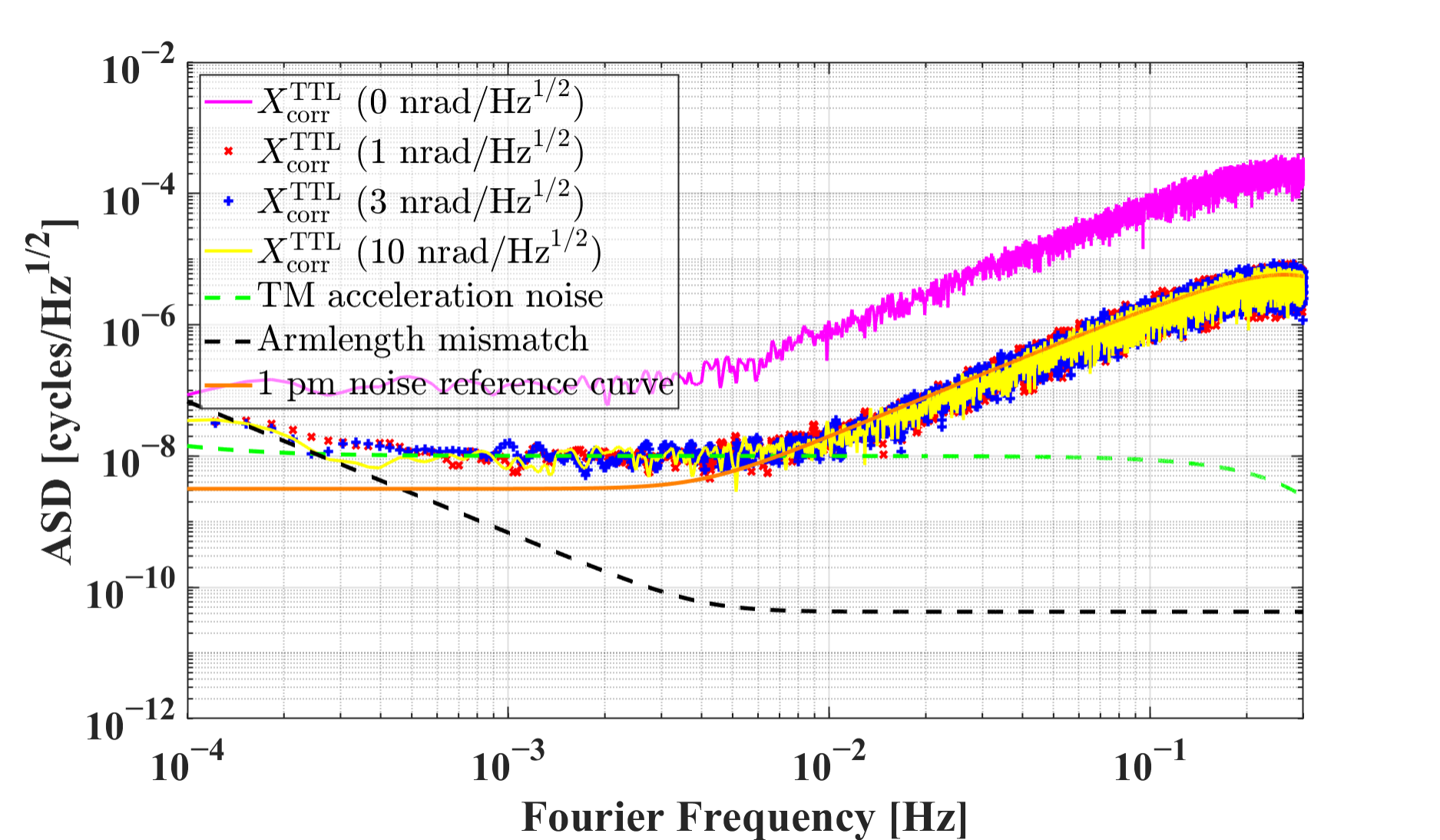}
\caption{\label{X_TTLc_angz} Residual noise levels in $X$ after removing the TTL noise and clock noise under different MOSA yaw jitter levels.}
\end{figure}

We also compute the residual noise levels in $X$ after clock noise removal and TTL noise correction with the estimated coefficients. As shown in Fig.~\ref{X_TTLc_angz}, we are unable to reduce the TTL noise in the case of zero MOSA yaw jitters (cyan) while the residual TTL noise in the other cases is no longer the dominant factor. The green and black traces are the same as in Fig.~\ref{X_TTLc}.

Based on the discussion, we find that higher MOSA yaw jitter levels would be benefit to the TTL noise estimation and subtraction. This reminds us to reconsider the filter passband we use in the simulations where we have assumed all SC and MOSA angular jitters have the same noise shape in the frequency domain. In a more realistic scenario, the SC and MOSA angular jitters may dominate in different frequency ranges~\cite{George2023} so that one may achieve better estimation accuracy of the TTL coefficients by choosing the frequency range that is dominated by the MOSA jitter.

\section{\label{sec7}Conclusion}

TTL noise is expected to be a significant limiting factor in TianQin's laser interferometers. Previous studies have introduced a solution for LISA to estimate and subtract the TTL noise with the Michelson TDI combination. In this article, we show as an alternative that TTL coefficients with linear drifts are estimated using the null TDI channel $\zeta$. We introduced the TTL noise model in TDI and the approach to noise estimation and subtraction, which is validated by our numerical simulation results. We generated a dataset of one-day period with an up-to-date instrumental noise setup. The estimation errors of all 24 linear TTL coefficients at time $t_0$ and 24 linear drift coefficients are within $10\ \upmu\text{m/rad}$ and $10\ \upmu\text{m/rad/day}$ respectively. The estimated coefficients are then used to subtract the TTL noise in the intermediary variable $\eta_i$. After TTL noise correction, one is able to construct the Michelson variables $X$, $Y$ and $Z$ for GW analysis.

Because of the point-ahead angles and the application of static compensation for PAAs in TianQin, TTL coefficients are slowly varying. We derive the expression of TTL noise induced by pointing errors using Zernike polynomials and find non-negligible coefficient drifts and quadratic TTL noise terms. The estimation error will below $0.3\ \text{pm/Hz}^{1/2}$ only if we take the linear drift coefficients and quadratic TTL coefficients into account. However, these coefficients depends highly on the actual wavefront characteristics. Therefore, more studies are required to determine whether these coefficients should be included in the estimation process. Nonetheless, we suggest taking the linear drift coefficients into account in postprocessing after the estimated linear TTL coefficients show long-term changes.

The MOSA yaw jitter level significantly affects the estimation error when using the null channel $\zeta$ in our simulations. The SC yaw jitter that is common to the two onboard MOSAs causes a great correlation, which accounts for the degradation of the estimated results in the absence of MOSA yaw jitter. In reality, the noise shapes of the SC and MOSA angular jitters may be different, indicating that changing the frequency range used for the fit may reduce potential correlation caused by the SC angular jitter and helps to increase estimation accuracy.

Note that all null TDI channels require six available links, for which we are unable to implement this approach with a lost arm. In this case, the Michelson combination is preferred. In addition, we would expect that glitches and data gaps with a long duration might greatly degrade the estimation performance, especially for the drift coefficients. In future work, we will consider the performance of the attitude and pointing control system for the SC and MOSA angular jitters used for TTL estimation. It is also an option that we translate the approach into a full Bayesian data analysis framework with unknown noise profile.

\begin{acknowledgments}
The authors would like to thank Bobing Ye for providing the TianQin orbit. The authors would also like to thank Jinmeng Wang for the useful discussions. This work was supported by the National Key R\&D Program of China (Grants No. 2020YFC2200202 and No. 2022YFC2204001).
\end{acknowledgments}

\appendix

\begin{widetext}
\section{\label{appA}Noise expressions and PSD models}

The clock noise estimate for $X$ reads
\begin{align}
	\hat{X}^q &= (a_{1'}f_1^{\text{u}}+a_3 f_3^{\text{u}})((2-{\bf D}_{31'}{\bf D}_{1'3}-{\bf D}_{31'}{\bf D}_{1'3}{\bf D}_{2'1}{\bf D}_{12'})(r_1+{\bf D}_{2'1}r_{2'})-(1-{\bf D}_{2'1}{\bf D}_{12'})(r_{1'}+{\bf D}_{31'}r_3)) \nonumber\\ &\quad - (a_1 f_1^{\text{u}}+a_{2'}f_2^{\text{u}})((2-{\bf D}_{2'1}{\bf D}_{12'}-{\bf D}_{2'1}{\bf D}_{12'}{\bf D}_{31'}{\bf D}_{1'3})(r_{1'}+{\bf D}_{31'}r_3)-(1-{\bf D}_{31'}{\bf D}_{1'3})(r_1+{\bf D}_{2'1}r_{2'})) \nonumber\\ &\quad + a_{2'}f_2^{\text{u}}(1-{\bf D}_{31'}{\bf D}_{1'3}-{\bf D}_{31'}{\bf D}_{1'3}{\bf D}_{2'1}{\bf D}_{12'}+{\bf D}_{2'1}{\bf D}_{12'}{\bf D}_{31'}{\bf D}_{1'3}{\bf D}_{31'}{\bf D}_{1'3})r_1 \nonumber\\ &\quad - a_3 f_3^{\text{u}}(1-{\bf D}_{2'1}{\bf D}_{12'}-{\bf D}_{2'1}{\bf D}_{12'}{\bf D}_{31'}{\bf D}_{1'3}+{\bf D}_{31'}{\bf D}_{1'3}{\bf D}_{2'1}{\bf D}_{12'}{\bf D}_{2'1}{\bf D}_{12'})r_{1'} \nonumber\\ &\quad + b_1 f_1^{\text{u}}(1-{\bf D}_{2'1}{\bf D}_{12'}-{\bf D}_{2'1}{\bf D}_{12'}{\bf D}_{31'}{\bf D}_{1'3}+{\bf D}_{31'}{\bf D}_{1'3}{\bf D}_{2'1}{\bf D}_{12'}{\bf D}_{2'1}{\bf D}_{12'})(r_{1'}+{\bf D}_{31'}r_3), \label{X^q}
\end{align}
where we have used that $b_{1'} = -b_1$. Eq.~(\ref{X^q}) is used for clock noise removal for $X$, i.e.,
\begin{equation}
	X_{\text{corr}}^q = X - \hat{X}^q. \label{X_corr^q}
\end{equation}

The clock noise estimate for $\zeta$ reads
\begin{align}
	\hat{\zeta}^q &= (a_{1'}-a_1-b_{1'})f_1^{\text{u}}(r_{3'}-r_3+{\bf D}_{23'}r_2+{\bf D}_{1'3}{\bf A}_{12'}(r_{2'}-r_2)-({\bf D}_{23'}-{\bf D}_{1'3}{\bf A}_{12'}){\bf D}_{3'2}{\bf A}_{31'}r_{1'}) \nonumber\\ &\quad + b_1 f_1^{\text{u}}({\bf D}_{23'}r_2-{\bf D}_{23'}{\bf D}_{3'2}{\bf A}_{31'}(r_{1'}-r_1)+(1-{\bf D}_{23'}{\bf D}_{3'2}{\bf A}_{31'}{\bf D}_{2'1}{\bf A}_{23'})(r_{3'}-r_3)) \nonumber\\ &\quad + (a_{2'}-b_{2'})f_2^{\text{u}}({\bf D}_{1'3}{\bf A}_{12'}r_2-{\bf D}_{1'3}{\bf A}_{12'}{\bf D}_{3'2}{\bf A}_{31'}(r_{1'}-r_1)) - a_2 f_2^{\text{u}}(r_{3'}-r_3+{\bf D}_{1'3}{\bf A}_{12'}r_{2'}) \nonumber\\ &\quad + b_2 f_2^{\text{u}}(r_{3'}-r_3+{\bf D}_{23'}r_2+{\bf D}_{1'3}{\bf A}_{12'}(r_{2'}-r_2)-({\bf D}_{23'}-{\bf D}_{1'3}{\bf A}_{12'}){\bf D}_{3'2}{\bf A}_{31'}(r_{1'}-r_1)) \nonumber\\ &\quad + (a_{3'}-a_3-b_{3'})f_3^{\text{u}}({\bf D}_{23'}r_2-{\bf D}_{23'}{\bf D}_{3'2}{\bf A}_{31'}(r_{1'}-r_1)+(1-{\bf D}_{23'}{\bf D}_{3'2}{\bf A}_{31'}{\bf D}_{2'1}{\bf A}_{23'})r_{3'}) \nonumber\\ &\quad + b_3 f_3^{\text{u}}(r_{3'}-r_3+{\bf D}_{23'}r_2+{\bf D}_{1'3}{\bf A}_{12'}(r_{2'}-r_2)). \label{zeta^q}
\end{align}
Eq.~(\ref{zeta^q}) is used for clock noise removal for $\zeta$, i.e.,
\begin{equation}
	\zeta_{\text{corr}}^q = \zeta - \hat{\zeta}^q. \label{zeta_corr^q}
\end{equation}

The PSD model of TM acceleration noise in $X$ reads
\begin{equation}
	S_{\text{TM}}^X(f) = \frac{16}{\lambda^2(2\pi f)^4}\sin^2(4\pi f d_0)\sin^2(2\pi f d_0)\left(8\cos^2(2\pi f d_0)+8\right)S_{\text{TM}}(f). \label{PSD_X_TM}
\end{equation}

The PSD model of sideband readout noise in $\zeta$ reads
\begin{equation}
	S_{\text{ro,sb}}^{\zeta}(f) = \frac{1}{k^2\lambda^2}M_{\text{ro,sb}}^{\text{isi}}S_{\text{ro,sb}}^{\text{isi}}(f) + \frac{1}{2k^2\lambda^2}M_{\text{ro,sb}}^{\text{rfi}}S_{\text{ro,sb}}^{\text{rfi}}(f) \label{PSD_zeta_ro_sb}
\end{equation}
where
\begin{align}
	M_{\text{ro,sb}}^{\text{isi}} &= \left|(a_{2'}-b_{2'}-b_2)+(b_1+b_2+a_{3'}-a_3-b_{3'})\text{e}^{-j\omega d_0}\right|^2 \nonumber\\ &\quad + \left|(a_{1'}-a_1-b_{1'}-a_{2'}+b_{2'}+b_2)-(a_{1'}-a_1-b_{1'}+b_1+b_2+a_{3'}-a_3-b_{3'})\text{e}^{-j\omega d_0}\right|^2 \nonumber\\ &\quad + \left|(a_{2'}-a_{1'}+a_1+b_{1'}-b_{2'}-b_2-b_3)+(a_{1'}-a_1-b_{1'}+b_1+b_2+a_{3'}-a_3-b_{3'}+b_3)\text{e}^{-j\omega d_0}\right|^2 \nonumber\\ &\quad + \left|a_{1'}-a_1-b_{1'}+b_2-a_2+b_3\right|^2 + \left|(a_{2'}-a_{1'}+a_1+b_{1'}-b_1-b_2-b_3)+b_1\text{e}^{-j\omega d_0}\right|^2 \nonumber\\ &\quad + \left|(a_{1'}-a_1-b_{1'}+b_1+b_2-a_2+a_{3'}-a_3-b_{3'}+b_3)-(b_1+a_{3'}-a_3-b_{3'})\text{e}^{-j\omega d_0}\right|^2
\end{align}
and
\begin{align}
	M_{\text{ro,sb}}^{\text{rfi}} &= \left|(a_{2'}-a_{1'}+a_1+b_{1'}-b_{2'}-b_2)+(a_2+a_{3'}-a_3-b_{3'}-b_3)\text{e}^{-j\omega d_0}+b_1\text{e}^{-2j\omega d_0}\right|^2 \nonumber\\ &\quad + \left|(a_2-a_{1'}+a_1+b_{1'}-b_2-b_3)+(a_{2'}-b_{2'}-b_2)\text{e}^{-j\omega d_0}+(a_{3'}-a_3+b_1+b_2-b_{3'})\text{e}^{-2j\omega d_0}\right|^2 \nonumber\\ &\quad + \left|\begin{array}{l}(a_2-a_{1'}+a_1+b_{1'}-b_1-b_2-a_{3'}+a_3+b_{3'}-b_3)\\+(a_{2'}+a_{3'}-a_3-a_{1'}+a_1+b_{1'}+b_1-b_2-b_{2'}-b_{3'}-b_3)\text{e}^{-j\omega d_0}\\+(a_{1'}-a_1-b_{1'}+b_1+b_2+a_{3'}-a_3-b_{3'}+b_3)\text{e}^{-2j\omega d_0}\end{array}\right|^2
\end{align}
with $\omega = 2\pi f$. The value of $S_{\text{ro,sb}}^{\zeta}$ in Eq.~(\ref{PSD_zeta_ro_sb}) will be reduced by half if we make use of both upper sideband and lower sideband measurements.
\end{widetext}

\bibliography{library}

\end{document}